\documentclass[a4paper,11pt]{article}
\usepackage{jcappub}
\usepackage{lineno}
\usepackage[compat=1.1.0]{tikz-feynman}
\usepackage{braket}
\usepackage{tikz}
\usetikzlibrary{arrows.meta}
\usepackage{orcidlink}
\usepackage{subcaption}
\usetikzlibrary{calc,intersections}
\usetikzlibrary{
    arrows.meta,
    decorations.markings,
    calc,
    angles,
    quotes
}
\usepackage{tikz-3dplot}
\newcommand{\pscale}{1.35}
\newcommand{\ball}[4]{%
  \node[
      circle,
      ball color=#3,
      opacity=#4,
      minimum size={2*#2*\pscale cm},
      inner sep=0pt
  ]
  at (#1) {};
}
\newcommand{\ballring}[4]{%
  \node[
      circle,
      draw=#3,
      dashed,
      line width=0.3pt,
      opacity=#4,
      minimum size={2*#2*\pscale cm},
      inner sep=0pt
  ]
  at (#1) {};
}
\title{Gravitational lensing of neutrinos through the lens of decoherence}
\author{Bikash Kumar Acharya\,\orcidlink{0009-0001-0982-9144},}
\author{Ujjal Kumar Dey\,\orcidlink{0000-0002-9620-7561}}

\affiliation{Department of Physical Sciences, Indian Institute of Science Education and Research
Berhampur,\\Ganjam, Odisha, 760003, India}

\emailAdd{bikashka22@iiserbpr.ac.in}
\emailAdd{ujjal@iiserbpr.ac.in}

\abstract{
We propose a new source of neutrino decoherence arising from stochastic gravitational lensing by unresolved mass substructure. Unlike traditional models that treat the lensing potential as smooth and deterministic, we decompose the lensing potential into a smooth component and a stochastic population of compact clumps. The substructure induced fluctuations of the lensing potential generate stochastic corrections to the neutrino time delay, thereby producing random relative phases between different neutrino mass eigenstates. Averaging over the clump ensemble converts these phase fluctuations into an effective dampening of flavour coherence. This framework connects stochastic gravitational lensing with neutrino phenomenology and offers a novel route to probe dark matter substructure through astrophysical neutrino flavour transitions.
}

\begin{document}
\maketitle

\section{Introduction}
\label{sec:intro}
One of the major challenges in fundamental physics is reconciling quantum theory with gravitation, specifically in understanding space-time fluctuations in extreme environments. Neutrinos are fundamental particles in the Standard Model which can potentially offer a unique opportunity to investigate this structure. The discovery of neutrino oscillation~\cite{Kajita:2016cak, McDonald:2016ixn} more than two decades ago confirmed that neutrinos are not massless which opened the window to a plethora of new investigations in elementary particle physics. Recent advancements have enhanced our understanding of neutrino oscillation and related measurements as it has entered to a precision era. Still there are a few open questions like absolute neutrino masses, hierarchy in mass orderings (normal or inverted), the octant issue, CP violation in neutrinos etc. Current and forthcoming neutrino detectors and experiments~\cite{DUNE:2020fgq, Hyper-Kamiokande:2018ofw, T2K:2011qtm, IceCube:2013low, IceCube:2013cdw, IceCube:2002eys, IceCube-Gen2:2020qha, KM3NeT:2024jji, ANITA:2010hzc, GRAND:2018iaj, RNO-G:2023kag, Krizmanic:2019hiq, ARIANNA:2014fsk} are expected to shed light on a number of such unanswered questions. In recent times, neutrino oscillations in curved space-time have gained significant attention~\cite{Cardall:1996cd, Fornengo:1996ef, Shi:2025plr, Shi:2025rfq, Shi:2024flw, Volpe:2023met, Lambiase:2022ucu, Buoninfante:2023qbk, Lambiase:2023pxd, AraujoFilho:2024mvz}. The interest arises from the fact that such analysis is not only sensitive to the underlying geometry of space-time which makes it relevant to different theories of gravity~\cite{Ahluwalia:1996ev, Grossman:1996eh, Geralico:2012zt, Luongo:2011zza, Buoninfante:2019der}, but also reveals unique characteristics of the neutrino sector that are absent in flat space-time. One major implication of such study is the increase in the oscillation length of neutrinos when influenced by curvature. Additionally, phenomena such as spin-flip or helicity transitions~\cite{Sorge:2007zza, Lambiase:2005gt} and potential violations of the equivalence principle~\cite{Lambiase:2001jr, Bhattacharya:1999na} have been explored within this gravitational framework. These effects illustrate that neutrinos, because of their weak interactions and long coherence lengths can retain information about gravitational environments that may be inaccessible through electromagnetic and gravitational wave observations alone.
Neutrinos propagating over astronomical distances experience loss of coherence from multiple microscopic and macroscopic effects~\cite{Acharya:2025rnw, DeRomeri:2023dht}. Several approaches to quantum gravity suggest that space-time may possess an intrinsically fluctuating structure which can induce stochastic uncertainties in distances, propagation times and particle dispersion relations~\cite{Ng:1993jb, Ng:2000fq, Ng:1999hm, Susskind:1994vu, Gambini:2006ph, Ng:2003jk}. In flat space-time, decoherence induced by quantum gravity motivated modified dispersion relations (MDR) damps oscillations proportional to the mass squared splittings. Separately, weak gravitational lensing of neutrinos combined with stochastic fluctuations in the phase around the gravitating lens introduces an additional decoherence channel whose strength depends on the geometrical parameters of the gravitating lens. A particularly interesting gravitational environment is provided by lensing systems. In the standard description of gravitational lensing of light, a smooth lensing potential produces multiple images, magnifications and time delays. For neutrinos, such lensing effects can modify the relative phases of the propagating mass eigenstates and may lead to interference between different classical paths. However, realistic lenses are not perfectly smooth. Galaxies and clusters contain unresolved dark matter substructure, compact clumps and small scale inhomogeneities. The lensing potential can therefore be decomposed into a smooth component and a stochastic substructure component. The additional stochastic component can perturb the lensing time delay and consequently inducing corrections to the neutrino oscillation phase.
In the presence of unresolved lens substructure, neutrino will no longer be propagating through a  smooth gravitational background. Instead, the presence of small scale clumps in the lens generate stochastic perturbations to the lensing geometry resulting in modifications to the path length and corresponding time delay. Hence, different realizations of the substructure will lead to slightly different oscillation phases. Since, the detailed configuration of the clumps may not be directly resolved by present day observations, the physically relevant oscillation probability should be understood as an averaged quantity over the possible substructure realizations. This averaging has an important consequence. The coherent oscillation phase is no longer perfectly sharp, rather it acquires a statistical spread due to the fluctuating lensing time delay. The average phase shift only modifies the effective oscillation phase whereas the spread around this average suppresses the interference between different neutrino mass eigenstates. Therefore, the stochastic lensing environment acts as an effective source of decoherence for neutrino flavour oscillations. The lensing induced decoherence should be interpreted as an ensemble averaged loss of phase coherence caused by unresolved gravitational substructure rather than as a fundamental non-unitary effect at the microscopic level.
In this work, we propose that stochastic gravitational lensing by unresolved substructure provides a new mechanism for neutrino decoherence. The noteworthy point is that the substructure induced time delay fluctuation generates a relative phase between neutrino mass eigenstates. In the Gaussian phase averaging approximation, the unresolved substructure is assumed to generate a statistically distributed phase shift whose dominant effect is captured by its variance. The ensemble average over these random phases then suppresses the coherent interference terms producing an effective lensing induced damping of neutrino flavour oscillations. The stochastic lensing contribution acts as a geometrically induced environmental noise source for the neutrino subsystem. After ensemble averaging over the unresolved lens substructure, the oscillation probabilities acquire additional damping terms beyond the standard coherent oscillation phase. This allows us to define a lensing induced decoherence parameter $\Gamma_{ij}^{\rm lens}(E,L)$, whose structure is determined by the phase variance accumulated along the lensed trajectory. Unlike phenomenological decoherence models in which the damping scale is motivated by power law approximation, here the decoherence parameter is connected to lensing observables and the statistical distribution of substructure. 
From the observational point of view, the most promising arena is provided by high energy astrophysical neutrino telescopes, where neutrinos travel cosmological baselines and may originate from strongly lensed transient or steady sources. Experiments such as IceCube~\cite{IceCube:2002eys}, KM3NeT~\cite{KM3NeT:2024jji}, IceCube-Gen2~\cite{IceCube-Gen2:2020qha}, GRAND~\cite{GRAND:2018iaj}, RNO-G~\cite{RNO-G:2023kag}, POEMMA~\cite{Olinto:2023vmx} and other future facilities are expected to improve the reconstruction of neutrino energy, arrival direction, flavour composition and possible time correlated multi-messenger counterparts. In such systems, stochastic lensing induced decoherence would not necessarily appear as a sharp event by event signature but rather as a statistical modification of flavour ratios, coherence patterns and directional dependent damping in neutrino observables. Therefore, although the present work is largely theoretical, it provides a phenomenological framework through which future high energy neutrino observations can be used to constrain or benchmark the effect of unresolved gravitational substructure on neutrino flavour propagation. 
This article is structured as follows. In Sec.~\ref{sec:nuosc_weak_lensing}, we briefly review neutrino oscillations in the presence of weak lensing. In Sec.~\ref{sec:lensing_review}, we outline theoretical framework for weak gravitational lensing of neutrinos and go on to calculate lensed neutrino oscillation probabilities. In Sec.~\ref{sec:nu_phase} and  Sec.~\ref{sec:time_delay}, we discuss the neutrino phase evolution in stochastic substructure and the corresponding arrival time delay, respectively and go on to derive the decoherence strength. We present our results and discuss in Sec.~\ref{sec:resanddisc} before we summarise and conclude in Sec.~\ref{sec:sumandconc}.
 
\section{Gravitationally Lensed Neutrino Oscillations}
\label{sec:nuosc_weak_lensing}
We begin by briefly reviewing neutrino flavour evolution in weak gravitational field aiming to establish our notations and conventions required for the stochastic lensing framework developed in later sections. Using the Pontecorvo-Maki-Nakagawa-Sakata (PMNS) mixing matrix, neutrino flavour eigenstates can be expressed in terms of the neutrino mass eigenstates as
\begin{equation}
    |\nu_\alpha\rangle = \sum_{a=1}^{n} U^*_{\alpha a} |\nu_a\rangle,
    \label{eq:flavour_to_mass}
\end{equation}
with $n$ being the number of neutrino mass eigenstates and $U$ is the PMNS matrix. For the standard three-neutrino case, $n = 3$ and $\alpha = e, \mu, \tau$ for flavour eigenstates which participate in weak interaction. The unitary $n \times n$ PMNS matrix which relates the flavour states $|\nu_\alpha\rangle$ ($\alpha = e, \mu, \tau, \ldots$) to mass eigenstates $|\nu_a\rangle$ ($a = 1, 2, 3,\ldots, n$) is given by,
\begin{equation}
U = \begin{pmatrix}
U_{e1} & U_{e2} & U_{e3} & \ldots \\
U_{\mu 1} & U_{\mu 2} & U_{\mu 3} & \ldots \\
U_{\tau 1} & U_{\tau 2} & U_{\tau 3} & \ldots \\
\vdots & \vdots & \vdots & \ddots
\end{pmatrix}.
\label{eq:pmns_mat}
\end{equation}
Assuming neutrinos are produced initially in the flavour eigenstate $|\nu_\alpha\rangle$ at source ($S$) and then after travelling to the detector ($D$), the oscillation probability of $\nu_\alpha \to \nu_\beta$ at the detection point is given as
\begin{equation}
P_{\nu_\alpha \rightarrow \nu_\beta} = \left|\langle \nu_\beta| \nu_\alpha (t_D,x_D)\rangle \right|^2 = \sum_{i,j}U_{\beta i} U^*_{\beta j} U_{\alpha j} U^*_{\alpha i}\, \exp(-i(\Phi_i - \Phi_j))\,,
\label{eq:gen_prob}
\end{equation} 
where ($\Phi_i - \Phi_j$) is the relative phase between ($i,j$) neutrino mass eigenstates pair. In flat space-time, the phase accumulated by a neutrino mass eigenstate which can be approximated as a plane wave is given as~\cite{Akhmedov:2009rb, Akhmedov:2010ua},
\begin{equation}
\Phi_i = E_i \Delta t -  p_i \Delta x =  E_i(t_D-t_S) - p_i\cdot(x_D-x_S)\,.
\label{eq:nu_phase1}
\end{equation}
Since, neutrinos are ultra relativistic in nature ($E_i \gg m_i$) and considering that all the mass eigenstates in a flavour eigenstate initially produced at the source with equal momentum or energy, the phase difference can be obtained as~\cite{Akhmedov:2009rb, Akhmedov:2010ua}, 
\begin{equation}
\Delta \Phi_{ij} = \Phi_i-\Phi_j \simeq \frac{\Delta m_{ij}^2}{2 E}\, |x_D-x_S|,
\label{eq:nu_phase2}
\end{equation}
where $\Delta m_{ij}^2 = m_i^2 - m_j^2$ and $E$ is the neutrino energy. The flavour oscillation probability $P_{\nu_\alpha \rightarrow \nu_\beta}$ therefore depends on the mass squared splitting and not on the absolute masses of the neutrinos in this case. Substituting Eq.~\eqref{eq:nu_phase2} in Eq.~\eqref{eq:gen_prob} leads to the neutrino oscillation probability. In three flavour paradigm, the oscillation probability between neutrino flavours, $\nu_\alpha$ and $\nu_\beta$ is given by,
\begin{align}
P_{\nu_\alpha \rightarrow \nu_\beta}
=
\delta_{\alpha\beta}
&-4\sum_{i>j}
\operatorname{Re}
\left(
U^*_{\alpha i}U_{\beta i}
U_{\alpha j}U^*_{\beta j}
\right)
\sin^2\left(
\frac{\Delta m^2_{ij}L}{4E}
\right)
\nonumber\\
&+2\sum_{i>j}
\operatorname{Im}
\left(
U^*_{\alpha i}U_{\beta i}
U_{\alpha j}U^*_{\beta j}
\right)
\sin\left(
\frac{\Delta m^2_{ij}L}{2E}
\right),
\end{align}
where $L=|x_D-x_S|$ is the neutrino propagation length and $\delta_{\alpha\beta}$ is the Kronecker delta. 
The influence of space-time curvature on neutrino propagation has been studied in the literature \cite{Cardall:1996cd, Fornengo:1996ef}. In a gravitational background with curved space-time, the phase in Eq.~\eqref{eq:nu_phase1} is expressed in the covariant form and written as,
\begin{equation} 
\Phi_i = \int_S^D p_\mu^{(i)} \, dx^\mu,
\label{eq:nu_phase3}
\end{equation}
where $ p_\mu^{(i)} $ is the covariant four momentum associated with $i$-th neutrino mass eigenstate along the propagation along trajectory. For a massive particle, we can have
\begin{equation} 
p_\mu^{(i)} = m_i \, g_{\mu\nu} \, \frac{dx^\nu}{ds},
\label{eq:canonical_mom}
\end{equation}
with $ g_{\mu\nu} $ is the space-time metric and $ ds $ is the corresponding line element. By computing the phase $ \Phi_i $ for a specific gravitational background and concerned neutrino trajectory and then substituting into Eq.~\eqref{eq:gen_prob} the modified oscillation probability can be evaluated. As an illustrative example, this formalism has been applied to the space-time of a static and spherically symmetric object described by the Schwarzschild metric \cite{Cardall:1996cd,Fornengo:1996ef}. In standard coordinates, the line element can be given as
\begin{equation} 
ds^2 = B(r)\, dt^2 - \frac{1}{B(r)}\, dr^2 - r^2 d\theta^2 - r^2 \sin^2\theta \, d\phi^2,
\label{eq:schwarzschild}
\end{equation}
where $ B(r) = 1 - 2GM/r = 1 - R_s/r $. Here $ G $ is Newton’s constant, $ M $, the mass of the gravitating body and $ R_s $ its Schwarzschild radius. Since, Schwarzschild space-time is spherically symmetric, the motion can be confined to the equatorial plane $ \theta = \pi/2 $ where the gravitational field is isotropic. Then,  oscillation phase accumulated by the $ j $-th neutrino mass eigenstate $ |\nu_j\rangle $ over the baseline is given by,
\begin{equation} 
\Phi_j = \int_S^D p_\mu^{(j)} \, dx^\mu = \int_S^D \bigl(p_t dt + p_r dr + p_{\phi} d\phi) = \int_S^D \bigl( E_j(r)\, dt - p_j(r)\, dr - J_j(r)\, d\phi \bigr),
\label{eq:nu_phase4}
\end{equation}
where $E_j$ and $J_j$ are the conserved energy and angular momentum of $j$-th neutrino mass eigenstate, respectively. At this point, we emphasize that although neutrino flavour evolution is intrinsically quantum mechanical, the propagation phase in a sufficiently weak gravitational field may be evaluated along the corresponding classical trajectory within the eikonal approximation~\cite{Cardall:1996cd, Lipkin:2000mz, Swami:2020qdi}. We restrict ourselves throughout to the ultra-relativistic and weak lensing regime, for which this description is appropriate. For a non-radial trajectory characterized by an impact parameter $b$, the weak field result obtained in the limits $GM\ll r_S,r_D$ and $b\ll r_S,r_D$ is~\cite{Fornengo:1996ef,Swami:2020qdi},
\begin{equation}
\Phi_j(b)
\simeq
\frac{m_i^{2}}{2E}\,
(r_S+r_D)
\left[
1
+
\frac{b^{2}}{2r_Sr_D}
+
\frac{2GM}{r_S+r_D}
\right].
\label{eq:nu_phase6}
\end{equation}
Here, $r_S$ and $r_D$ denote the radial coordinates of the source and detector respectively. The precise form of the phase depends on the chosen trajectory and on the order retained in the weak field expansion.
From Eq.~\eqref{eq:nu_phase6}, it can be seen that the phase difference $ \Delta \Phi_{jk} $ depends only on $ \Delta m^2_{jk} $. Hence, even in curved space-time, as long as there is a single classical trajectory from $ S $ to $ D $, the oscillation probability depends only on $ \Delta m^2_{jk} $ as in flat space-time. However, we discuss in later sections that for propagation along a single classical trajectory, the oscillation phase remains controlled by the mass squared splitting even in a weak gravitational background. Gravitational lensing becomes relevant when the propagation geometry contains multiple paths or when the lensing potential contains unresolved stochastic substructure. In that case, small perturbations of the lensing time delay generate corresponding fluctuations in the neutrino oscillation phase. This observation provides the starting point for the stochastic lensing decoherence mechanism developed in the following sections.
%

\section{A Quick Look at Neutrino Gravitational Lensing}
\label{sec:lensing_review}
\begin{figure}[!ht]
\centering


\tdplotsetmaincoords{46}{-35}

\begin{tikzpicture}[
    tdplot_main_coords,
    scale=\pscale,
    >=stealth,
    axis/.style={thick,-{Stealth[length=2.5mm]}},
    ray/.style={
        black,
        line width=0.9pt,
        postaction={decorate},
        decoration={
            markings,
            mark=at position 0.72 with
            {\arrow{Stealth[length=2.5mm,width=1.8mm]}}
        }
    }
]


\coordinate (L) at (0,0,0);
\coordinate (S) at (-4.5,0,0);
\coordinate (O) at (4,1,0);

\coordinate (B1) at (0,1.15,0);
\coordinate (B2) at (0,-1.05,0);


\shade[
    top color=gray!8,
    bottom color=gray!18,
    opacity=0.55
]
(-5.6,-1.70,0)
--
(5.6,-1.55,0)
--
(5.6,1.55,0)
--
(-5.6,1.55,0)
-- cycle;

\draw[
    gray!55,
    line width=0.3pt,
    opacity=0.7
]
(-5.6,-1.55,0)
--
(5.6,-1.55,0)
--
(5.6,1.55,0)
--
(-5.6,1.55,0)
-- cycle;

\foreach \gx in {-5,-4,...,5}{
    \draw[
        gray!35,
        line width=0.15pt,
        opacity=0.5
    ]
    (\gx,-1.55,0) -- (\gx,1.55,0);
}

\foreach \gy in {-1.5,-1,-0.5,0,0.5,1,1.5}{
    \draw[
        gray!35,
        line width=0.15pt,
        opacity=0.5
    ]
    (-5.6,\gy,0) -- (5.6,\gy,0);
}


\begin{scope}[canvas is xy plane at z=0]

    \fill[
        orange!35!yellow,
        opacity=0.38
    ]
    (0,0) circle (1.55);

    \draw[
        orange!70!red,
        line width=0.8pt,
        opacity=1
    ]
    (0,0) circle (1.55);

\end{scope}


\ball{L}{1.55}{gray!35}{0.55}

\ballring{L}{1.10}{gray!70}{0.7}
\ballring{L}{0.65}{gray!70}{0.7}


\foreach \x/\y/\z/\rad in {
	-1.04/0.78/0.40/0.10,
    -0.45/1.02/-0.36/0.14,
     0.40/0.94/0.47/0.12,
     1.04/0.66/-0.29/0.09,
     -1.12/0.16/-0.45/0.08,
    -0.63/0.30/0.35/0.12,
     0.45/0.27/-0.40/0.15,
     1.09/0.13/0.29/0.08,
     -0.94/-0.66/0.36/0.11,
    -0.30/-0.97/-0.32/0.13,
     0.52/-0.83/0.37/0.10,
     1.07/-0.56/-0.25/0.08,
     -1.20/-0.29/0.18/0.08,
     1.15/-0.15/-0.22/0.07,
     0.11/0.04/0.62/0.09,
    -0.25/0.54/-0.58/0.08,
     0.68/-0.22/0.52/0.07
}{
    \ball{\x,\y,\z}{\rad}{gray!75}{0.95}
}


\ball{S}{0.60}{orange!30}{0.20}
\ball{S}{0.25}{orange!70!red}{1}
\ball{S}{0.09}{yellow!80}{1}


\ball{L}{0.28}{teal!55!green}{1}
\ball{L}{0.13}{teal!15!white}{0.9}


\ball{O}{0.55}{blue!25}{0.20}
\ball{O}{0.26}{blue!65!cyan}{1}
\ball{O}{0.10}{blue!15!white}{1}

\draw[
    black,
    line width=0.8pt
]
(S) -- (L) -- (O);



\draw[
    black,
    line width=0.9pt,
    postaction={decorate},
    decoration={
        markings,
        mark=at position 0.76 with
        {\arrow{Stealth[length=2.4mm,width=1.7mm]}}
    }
]
(S)
.. controls (-1.2,1.96,0) and (1.6,1.88,0)
.. (O)
coordinate[pos=0.48] (Bimp1);


\draw[
    black,
    line width=0.9pt,
    postaction={decorate},
    decoration={
        markings,
        mark=at position 0.76 with
        {\arrow{Stealth[length=2.4mm,width=1.7mm]}}
    }
]
(S)
.. controls (-1.2,-2.22,0) and (1.6,-2.22,0)
.. (O)
coordinate[pos=0.48] (Bimp2);


\draw[
    blue,
    line width=0.8pt
]
(L) -- (Bimp1)
node[
    midway,
    left=-1pt
]
{$b_1$};

\draw[
    blue,
    line width=0.8pt
]
(L) -- (Bimp2)
node[
    midway,
    right=-1pt
]
{$b_2$};

\draw[axis]
(0,0,0) -- (5.4,0,0)
node[below right] {$x$};

\draw[axis]
(0,0,0) -- (0,3.4,0)
node[above] {$y$};

\draw[axis]
(0,0,0) -- (0,0,2.6)
node[above] {$z$};


\node[below]
at ($(L)+(-1,-2.2,0)$)
{Gravitating lens};

\node[below]
at ($(S)+(-0.5,-0.95,0)$)
{Source};

\node[right]
at ($(O)+(0.55,0.35,0)$)
{Observer};

\node[below]
at (-2.3,0.05,0)
{$r_S$};

\node[below]
at (2.2,0.5,0)
{$r_D$};
\end{tikzpicture}
\caption{Diagrammatic illustration of gravitational lensing of neutrinos in the equatorial plane. Here, the dominant gravitating lens is embedded within a smooth gravitational halo containing a stochastic population of unresolved substructures.}
\label{fig:diagram}
\end{figure}
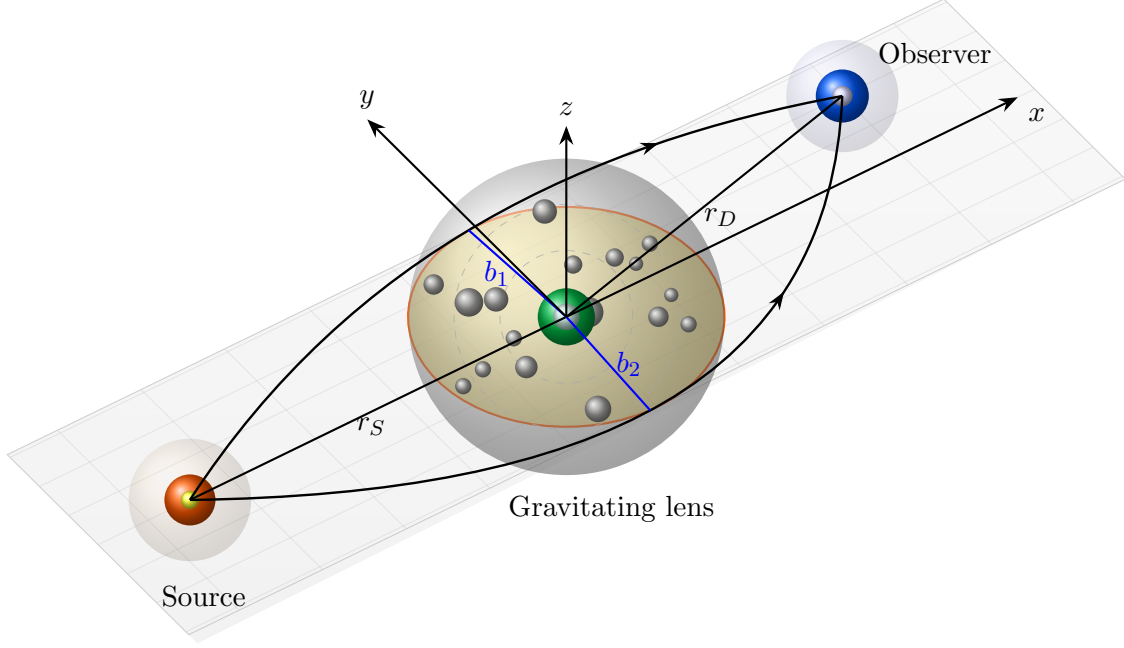

This section briefly summarizes the conventional picture of neutrino gravitational lensing in a smooth gravitational background. In such case, the gravitational field is assumed to be smooth and deterministic. Unlike propagation along a unique trajectory, gravitational lensing may allow more than one classical path to connect a neutrino source and an observer. The corresponding mass eigenstates can therefore interfere at the detector with each trajectory carrying a different geometrical propagation phase \cite{Fornengo:1996ef, Crocker:2003cw, Alexandre:2018crg, Swami:2020qdi}. This provides a useful reference point for distinguishing coherent multi path interference from the stochastic lensing mechanism considered later in this work. Non-radial propagation of neutrinos  around a gravitating lens, the dependence of their quantum phase on the impact parameter $ b $ leads to lensing effects. Considering a Schwarzschild black hole acting as a gravitational lens between a neutrino source and a detector, as illustrated in Fig.~\ref{fig:diagram}.
In such setting, neutrinos of a given mass eigenstate $ |\nu_i\rangle $ travel along distinct classical paths and interfere at a common detection point $ D $. The neutrino flavour arriving at the detector after propagating along multiple paths labeled as $ p $ is given as,
\begin{equation} 
|\nu_\alpha (t_D,x_D)\rangle = N \sum_{i} U_{\alpha i}^* \sum_p \exp \bigl(-i \Phi_i^p\bigr) \,|\nu_i (t_S,x_S)\rangle,
\label{eq:time_evolution_modified}
\end{equation}
where $\Phi_j^p$ is the phase in Eq.~\eqref{eq:nu_phase6} but now it has a path dependent (impact) parameter $b_p$. If the neutrino is produced with flavour $ \alpha $ at the source $ S $, the probability of detecting it in flavour $ \beta $ at $ D $ is given as,
\begin{equation} 
P^{\rm lens}_{\nu_\alpha \rightarrow \nu_\beta} = \bigl|\langle \nu_\beta | \nu_{\alpha}(t_D,x_D)\rangle\bigr|^2
= |N|^2 \sum_{i,j} U_{\beta i} U^*_{\beta j} U_{\alpha j} U^*_{\alpha i}
\sum_{p,q} \exp \Bigl(-i \Delta \Phi^{pq}_{ij}\Bigr),
\label{eq:gen_lensed_prob}
\end{equation}
with normalization constant
\begin{equation} 
|N|^2 = \Bigl( \sum_i |U_{\alpha i}|^2 \sum_{p,q}\exp \bigl(-i \Delta \Phi^{pq}_{ii}\bigr) \Bigr)^{-1},
\label{eq:norm}
\end{equation}
which ensures total probability conservation when multiple coherent paths contribute to the detected amplitude. In the weak field limit, using Eq.~\eqref{eq:nu_phase6}, the phase difference can be decomposed into a part that depends on the mass squared splitting $ \Delta m^2_{ij} $ and a part that depends on the path difference resulting from different impact parameters $ \Delta b_{pq}^2 $ given as
\begin{equation} 
\Delta\Phi_{ij}^{pq} = \Phi_i^p - \Phi_j^q
= \Delta m^2_{ij}\, A_{pq} + \Delta b_{pq}^2\, B_{ij},
\label{eq:phase_diffe}
\end{equation}
where the co-efficients can be evaluated as~\cite{Swami:2020qdi},
\begin{align}
A_{pq} & = \frac{r_S + r_D}{2E}
        \left[ 1 + \frac{2~GM}{r_S + r_D} - \frac{b_p^2 + b_q^2}{4 r_S r_D} \right], \label{eq:A_pq} \\
B_{ij} & = - \frac{m_i^2 + m_j^2}{8E}
        \left[ \frac{1}{r_S} + \frac{1}{r_D} \right]. \label{eq:B_ij}
\end{align}
At this point, we emphasize that the first term in Eq.~\eqref{eq:phase_diffe} has the trivial dependence on the neutrino mass squared splitting whereas the second term couples the difference between the classical trajectories to the combination $m_i^2+m_j^2$. Thus, the lensing geometry introduces phase information that is absent in ordinary single path propagation. In particular, when $\Delta b_{pq}^{2}=0$, the path difference contribution vanishes and the dependence on the absolute mass combination through $B_{ij}$ is lost. For the weak lensing configuration restricted to the equatorial plane,
two classical trajectories are relevant. Denoting their impact
parameters by $b_1$ and $b_2$ and defining $\Delta b^2 = b_1^2-b_2^2$, the lensed oscillation probability can be obtained as,
\begin{align} 
P_{\nu_\alpha \rightarrow \nu_\beta}^{\rm lens}
&= |N|^2 \Biggl[
   2\sum_{i} |U_{\beta i}|^2 |U_{\alpha i}|^2
   \bigl(1+\cos(\Delta b^2 B_{ii})\bigr)
   \nonumber \\
   &\; + \sum_{i,j\neq i} U_{\beta i} U^*_{\beta j} U_{\alpha j} U^*_{\alpha i}
   \Bigl( e^{-i\Delta m_{ij}^2 A_{11}} + e^{-i\Delta m_{ij}^2 A_{22}}
         + 2\cos \bigl(\Delta b^2 B_{ij}\bigr)
           e^{-i\Delta m_{ij}^2 A_{12}} \Bigr) \Biggr],
\label{eq:lensed_prob2}
\end{align}
with the normalization constant
\begin{equation} 
|N|^2 = \Bigl[2 + 2\sum_i |U_{\alpha i}|^2 \cos\!\bigl(\Delta b^2 B_{ii}\bigr) \Bigr]^{-1},
\label{eq:norm1}
\end{equation}
where $ \Delta b^2 = \Delta b_{12}^2 $. The two flavour consequences of Eq.~\eqref{eq:lensed_prob2} including the
dependence of the interference pattern on the individual neutrino
masses and on the mass ordering have been studied in detail in the
literature~\cite{Swami:2020qdi, Shi:2024flw, Shi:2025rfq, Shi:2025plr}. The rationale here is that we extract a compact three flavour form that can subsequently be generalized to a stochastic lensing environment. For a realistic three flavour paradigm, the lensed oscillation probability can be evaluated by following the method described in Ref.~\cite{Swami:2020qdi}. With a few new definitions, the relevant parameters can now be promoted with appropriate indices as below,
\begin{align}
    \Delta m^2_{ij} &= m_i^2 - m_j^2, & 
    \sum m^2_{ij}    &= m_i^2 + m_j^2, \\
    \xi_{ij} &= \frac{r_S+r_D}{8Er_Sr_D} \Delta b^2 \Delta m^2_{ij}, & 
    \epsilon_{ij} &= \Delta b^2 B_{ij} 
    = -\frac{\Delta b^2 \sum m^2_{ij}}{8E}\left(\frac{1}{r_S} + \frac{1}{r_D}\right).
\end{align} 
Now, we proceed with the small parameter expansion of Eq.~\eqref{eq:lensed_prob2} in $\Delta b^2$ and truncating it at the second order, we obtain the relevant probability terms. Demanding $ \sum_{\beta} P^{\rm lens}_{\nu_\alpha \to \nu_\beta} = 1 $, we obtain the normalization constant in this case as,
\begin{equation}
|N|^2 = \frac{1}{4}\left(1 + \frac{1}{4}\sum_i |U_{\alpha i}|^2 \epsilon_{ii}^2 + \mathcal{O}(\epsilon^4)\right).
\label{eq:3f_norm}
\end{equation}
The zeroth order term, which can be evaluated by setting $ \epsilon_{ij}=0 $ ($ \Delta b^2=0 $) in the expansion, is given as, 
\begin{align}
P^{\rm lens(0)}_{\nu_\alpha \to \nu_\beta} 
&= \delta_{\alpha\beta} 
   - 4\sum_{i<j} \operatorname{Re}(U_{\alpha i}^* U_{\beta i} U_{\alpha j} U_{\beta j}^*) \sin^2\left(\frac{\Delta m^2_{ij} A_{12}}{2}\right)
   \nonumber \\
   &\quad + 2\sum_{i<j} \operatorname{Im}(U_{\alpha i}^* U_{\beta i} U_{\alpha j} U_{\beta j}^*) \sin\left(\Delta m^2_{ij} A_{12}\right),
\label{zeroth_term}
\end{align}
with $ A_{12} $ playing the role of $ L/2E $. This is exactly the standard three flavour oscillation probability. In line with this, the first non-trivial correction in the expansion comes from the second order term, which is given by,
\begin{equation}
\begin{aligned}
P^{\rm lens\,(2)}_{\nu_\alpha\to\nu_\beta}
&=
\left(
\frac{r_S+r_D}{8Er_Sr_D}\,
\Delta b^2
\right)^2
\Bigg[
P_{\nu_\alpha\rightarrow\nu_\beta}^{(0)}
\sum_{k=1}^{3}
|U_{\alpha k}|^2m_k^4
\\
&\hspace{0.5cm}
-
\sum_{i=1}^{3}
|U_{\alpha i}|^2|U_{\beta i}|^2m_i^4
\\
&\hspace{0.5cm}
-
\sum_{i<j}
\left(m_i^4+m_j^4\right)
\Bigg\{
\operatorname{Re}\!\left(
U_{\beta i}U_{\beta j}^{*}
U_{\alpha i}^{*}U_{\alpha j}
\right)
\cos\!\left(\Delta m_{ij}^{2}A_{12}\right)
\\
&\hspace{3.6cm}
+
\operatorname{Im}\!\left(
U_{\beta i}U_{\beta j}^{*}
U_{\alpha i}^{*}U_{\alpha j}
\right)
\sin\!\left(\Delta m_{ij}^{2}A_{12}\right)
\Bigg\}
\Bigg]
\\
&\hspace{0.5cm}
+
\mathcal{O}\!\left[
(\Delta b^2)^4
\right].
\end{aligned}
\label{eq:second_term}
\end{equation}
Hence, the total lensed oscillation probability up to $ \mathcal{O}\!\left[(\Delta b^2)^2\right] $, can be given as
\begin{equation}
P^{\rm lens}_{\nu_\alpha\to\nu_\beta}
=
P^{\rm lens (0)}_{\nu_\alpha\to\nu_\beta} + 
P^{\rm lens (2)}_{\nu_\alpha\to\nu_\beta}
\end{equation}
In a nut shell, the discussion in this section focuses on the coherent lensing of neutrinos in a smooth and deterministic gravitational background. In such scenario, different classical paths accumulates distinct phases and can interfere at the detector. In realistic lensing systems, the smooth gravitational potential is accompanied by unresolved substructure, whose detailed configuration is not known event by event. These small scale perturbations induce stochastic corrections to the lensing time delay and hence to the neutrino oscillation phase. In the following sections, we promote the deterministic lensing phase to a statistically fluctuating quantity and show how ensemble averaging over the substructure population leads to an effective decoherence of neutrino flavour oscillations.
%
\section{Phase Evolution in a Stochastic Lensing Geometry}
\label{sec:nu_phase}
The discussion in the previous section describes the accumulation of neutrino oscillation phases along prescribed classical trajectories. However, in a gravitational lensing geometry, the source and detector may be connected by more than one classical path. For a smooth and deterministic lens each path contributes a well defined propagation phase and the detected neutrino flavour state is obtained by summing the amplitudes associated with the different lensed trajectories. This multi path picture provides us the natural initiation point for understanding how gravitational lensing can modify neutrino flavour oscillations. Taking motivation from the prescription discussed in Ref.~\cite{Keeton:2009ua, DiazRivero:2017xkd, DiazRivero:2018oxk}, under stochastic lensing, the phase of the $i$-th neutrino mass eigenstate along a lensed trajectory $p$ can be written as,
\begin{equation}
\Phi_i^p = \Phi_i^{p,0} + \delta\Phi_i^{p,s},
\end{equation}
where $\Phi_i^{p,0}$ is the phase accumulated in the smooth lensing background whereas $\delta\Phi_i^{p,s}$ represents the perturbation induced by unresolved lens substructure. The smooth contribution is deterministic once the macroscopic lens model is assumed. In contrast, the substructure contribution depends on the detailed distribution of compact clumps and hence it is stochastic in nature. The neutrinos arriving at the detector can be written as a sum over mass eigenstates and lensing trajectories,
\begin{equation}
|\nu_\alpha(D)\rangle = N' \sum_i U_{\alpha i}^{*} \sum_p \exp\left[-i\left(\Phi_i^{p,0} + \delta\Phi_i^{p,s}\right)\right] |\nu_i\rangle ,
\end{equation}
where $N'$ represents the normalization factor with substructure effects. The corresponding lensed oscillation probability contains interference terms of the form
\begin{equation}
\exp\left[
-i\left(
\Phi_i^{p,0}
-
\Phi_j^{q,0}
\right)
\right]
\left\langle
\exp\left[
-i\left(
\delta\Phi_i^{p,s}
-
\delta\Phi_j^{q,s}
\right)
\right]
\right\rangle_{\rm sub},
\label{eq:modified_phase}
\end{equation}
where \(\langle\cdots\rangle_{\rm sub}\) represent an ensemble averaging over stochastic realizations of the unresolved substructure. As we can see, Eq.~\eqref{eq:modified_phase} clearly makes the distinction between smooth lensing and stochastic lensing. The smooth lens gives the deterministic phase difference between the classical paths whereas the unresolved substructure introduces a statistical spread in the phase. If the substructure induced phase fluctuations are assumed to be Gaussian~\cite{Ohlsson:2000mj, Petruzziello:2020wkd, Al-Nasrallah:2021zie} then the ensemble average suppresses the interference term through a factor parametrized by the phase variance and is given as
\begin{equation}
\left\langle
\exp\left[
-i\left(
\delta\Phi_i^{p,s}
-
\delta\Phi_j^{q,s}
\right)
\right]
\right\rangle_{\rm sub}
= 
\exp\left[
-\frac{1}{2}
\mathrm{Var}[\Delta\phi_{\rm sub}]
\right],
\end{equation}
where, we define the variance as
\begin{equation}
\mathrm{Var}[\Delta\phi_{\rm sub}]
=
\left\langle
\left(
\delta\Phi_i^{p,s}
-
\delta\Phi_j^{q,s}
\right)^2
\right\rangle_{\rm sub}
-
\left\langle
\delta\Phi_i^{p,s}
-
\delta\Phi_j^{q,s}
\right\rangle_{\rm sub}^{2}.
\label{eq:phase_variance}
\end{equation}
Thus, it can be seen that the smooth component produces path dependent phase shifts whereas the stochastic substructure component produces an effective damping of the interference terms after ensemble averaging. We identify the latter as stochastic lensing induced neutrino phase smearing (or decoherence). In the following section, we connect the stochastic phase perturbation $\delta\Phi_i^{p,s}$ to the substructure induced fluctuation of the lensing time delay and express the resulting decoherence scale in terms of lensing observables.

\section{Lensing Time Delay and Neutrino Phase Fluctuations}
\label{sec:time_delay}
In this section, we review the gravitational lensing time delay formalism~\cite{Coe:2009wt, Keeton:2009ua, Schneider:1992bmb, Narayan:1996ba, Penacchioni:2019czg, Penacchioni:2019kix} and establish the notation used later to describe stochastic lensing induced neutrino decoherence. The lensing potential $\phi(\vec x)$ is the scaled two dimensional Newtonian potential satisfying the relation,
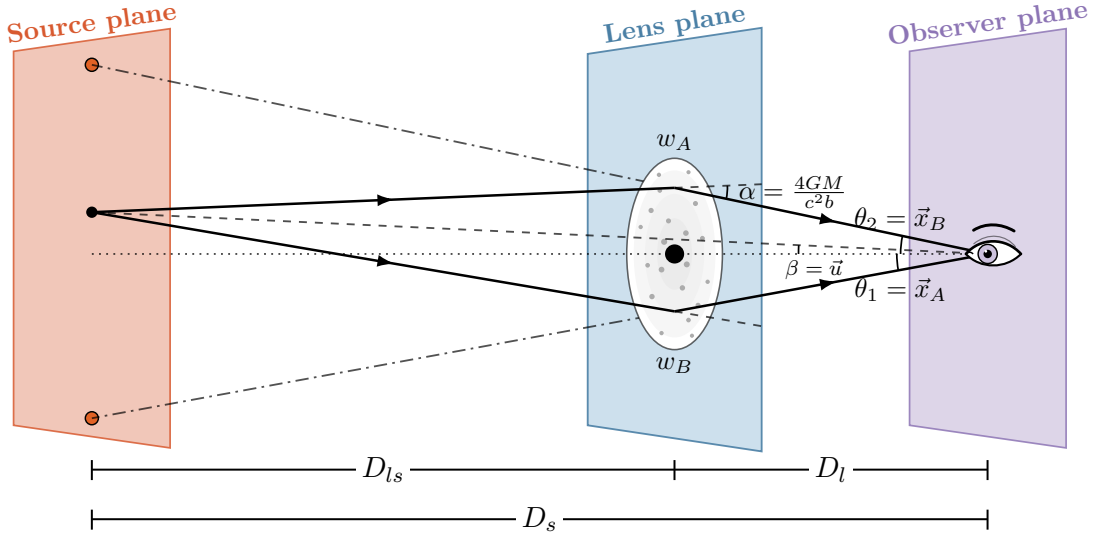
\begin{figure}
\definecolor{uldmblue}{HTML}{74A9CF}
\definecolor{uldmline}{HTML}{4A7FA5}
\definecolor{neutrino}{HTML}{D85A30}
\definecolor{scalarpurple}{HTML}{8D75B5}
\colorlet{sourceplane}{neutrino}
\colorlet{lensplane}{uldmblue}
\colorlet{observerplane}{scalarpurple}
\begin{tikzpicture}[
    scale=1.15,
    >=Latex,
    ray/.style={
        black,
        line width=0.95pt,
        postaction={decorate},
        decoration={
            markings,
            mark=at position 0.52 with
            {\arrow{Latex[length=2.3mm,width=1.6mm]}}
        }
    },
    construction/.style={
        black,
        dash pattern=on 5pt off 2pt on 1pt off 2pt,
        line width=0.70pt,
        opacity=0.68
    },
    dashedref/.style={
        black,
        dashed,
        line width=0.70pt,
        opacity=0.70
    },
    axisline/.style={
        black,
        dotted,
        line width=0.70pt,
        opacity=0.72
    },
    dimline/.style={
        black,
        line width=0.75pt
    },
    anglecurve/.style={
    draw=black,
    line width=0.65pt
	},
    callout/.style={
        black,
        line width=0.60pt,
        -{Latex[length=1.5mm,width=1.1mm]}
    }
]

\def\xS{0}
\def\xL{6.7}
\def\xO{10.3}

\coordinate (Saxis) at (\xS,0);
\coordinate (S) at (\xS,0.48);       
\coordinate (L) at (\xL,0);
\coordinate (WA) at (\xL,0.76);
\coordinate (WB) at (\xL,-0.66);
\coordinate (O) at (\xO,0);

\pgfmathsetmacro{\timage}{(\xO-\xS)/(\xO-\xL)}
\coordinate (IA) at ($(O)!\timage!(WA)$);
\coordinate (IB) at ($(O)!\timage!(WB)$);


\coordinate (SPL1) at ($(Saxis)+(-0.90,-1.97)$);
\coordinate (SPL2) at ($(Saxis)+( 0.90,-2.23)$);
\coordinate (SPL3) at ($(Saxis)+( 0.90, 2.59)$);
\coordinate (SPL4) at ($(Saxis)+(-0.90, 2.33)$);

\fill[sourceplane,opacity=0.34]
(SPL1)--(SPL2)--(SPL3)--(SPL4)--cycle;

\draw[sourceplane,opacity=0.85,line width=0.65pt]
(SPL1)--(SPL2)--(SPL3)--(SPL4)--cycle;

\node[
    text=sourceplane,
    font=\small,
    rotate=8
]
at ($(SPL4)!0.50!(SPL3)+(0,0.18)$)
{\textbf{Source plane}};

\coordinate (LPL1) at ($(L)+(-1.00,-1.97)$);
\coordinate (LPL2) at ($(L)+( 1.00,-2.27)$);
\coordinate (LPL3) at ($(L)+( 1.00, 2.60)$);
\coordinate (LPL4) at ($(L)+(-1.00, 2.30)$);

\fill[lensplane,opacity=0.36]
(LPL1)--(LPL2)--(LPL3)--(LPL4)--cycle;

\draw[uldmline,opacity=0.90,line width=0.70pt]
(LPL1)--(LPL2)--(LPL3)--(LPL4)--cycle;

\node[
    text=uldmline,
    font=\small,
    rotate=9
]
at ($(LPL4)!0.50!(LPL3)+(0,0.18)$)
{\textbf{Lens plane}};

\coordinate (OPL1) at ($(O)+(-0.90,-1.97)$);
\coordinate (OPL2) at ($(O)+( 0.90,-2.23)$);
\coordinate (OPL3) at ($(O)+( 0.90, 2.59)$);
\coordinate (OPL4) at ($(O)+(-0.90, 2.33)$);

\fill[observerplane,opacity=0.30]
(OPL1)--(OPL2)--(OPL3)--(OPL4)--cycle;

\draw[observerplane,opacity=0.85,line width=0.65pt]
(OPL1)--(OPL2)--(OPL3)--(OPL4)--cycle;

\node[
    text=observerplane,
    font=\small,
    rotate=8
]
at ($(OPL4)!0.50!(OPL3)+(0,0.18)$)
{\textbf{Observer plane}};


\filldraw[
    draw=black,
    line width=0.55pt,
    fill=sourceplane!78!orange
]
(IA) circle (0.075);

\filldraw[
    draw=black,
    line width=0.55pt,
    fill=sourceplane!78!orange
]
(IB) circle (0.075);

\fill[black] (S) circle (0.065);

\draw[construction] (WA)--(IA);
\draw[construction] (WB)--(IB);

\begin{scope}[shift={(L)}]

    \fill[white]
    (0,0) ellipse [x radius=0.55,y radius=1.10];

    \draw[
        black,
        opacity=0.60,
        line width=0.60pt
    ]
    (0,0) ellipse [x radius=0.55,y radius=1.10];

    \fill[gray!7]
    (0,0) ellipse [x radius=0.47,y radius=0.96];

    \fill[gray!11]
    (0,0) ellipse [x radius=0.34,y radius=0.69];

    \fill[gray!16]
    (0,0) ellipse [x radius=0.20,y radius=0.41];

    \begin{scope}
        \clip (0,0) ellipse [x radius=0.52,y radius=1.06];

        \foreach \x/\y/\r in {
            -0.12/ 0.14/0.035,
             0.13/ 0.24/0.032,
            -0.15/-0.18/0.033,
             0.14/-0.12/0.034,
            -0.07/ 0.38/0.029,
             0.09/-0.39/0.031,
            -0.27/ 0.50/0.030,
             0.24/ 0.58/0.029,
            -0.26/-0.49/0.031,
             0.27/-0.59/0.029,
            -0.17/ 0.73/0.027,
             0.16/-0.76/0.027,
             0.32/ 0.18/0.028,
            -0.31/-0.13/0.027,
            -0.18/ 0.91/0.025,
             0.16/ 0.94/0.024,
            -0.15/-0.91/0.025,
             0.20/-0.94/0.024,
            -0.39/ 0.31/0.025,
             0.38/-0.32/0.025
        }{
            \fill[gray!65] (\x,\y) circle (\r);
        }
    \end{scope}

    \fill[black] (0,0) circle (0.11);

\end{scope}


\draw[axisline] (Saxis)--(O);
\draw[dashedref] (S)--(O);


\draw[ray] (S)--(WA);
\draw[ray] (WA)--(O);
\draw[ray] (S)--(WB);
\draw[ray] (WB)--(O);


\node[
    font=\small
]
at ($(L)+(0,1.28)$)
{$w_A$};

\node[
    font=\small
]
at ($(L)+(0,-1.28)$)
{$w_B$};


\pgfmathsetmacro{\tincomingext}{(\xL+1.00-\xS)/(\xL-\xS)}
\coordinate (AincExt) at ($(S)!\tincomingext!(WA)$);
\coordinate (BincExt) at ($(S)!\tincomingext!(WB)$);

\draw[dashedref] (WA)--(AincExt);
\draw[dashedref] (WB)--(BincExt);

\pic[
    anglecurve,
    angle radius=7mm
]
{angle=O--WA--AincExt};

\node[
    font=\small,
    anchor=west
]
at ($(WA)+(0.62,-0.05)$)
{$\alpha = \frac{4GM}{c^2 b}$};


\pic[
    anglecurve,
    angle radius=11.5mm
]
{angle=WA--O--L};

\node[
    font=\small
]
at ($(O)+(-1,0.41)$)
{$\theta_2 = \vec x_B$};

\pic[
    anglecurve,
    angle radius=12mm
]
{angle=L--O--WB};

\node[
    font=\small
]
at ($(O)+(-1,-0.41)$)
{$\theta_1 = \vec x_A$};

\pic[
    anglecurve,
    angle radius=25mm
]
{angle=S--O--L};

\node[
    font=\scriptsize
]
at ($(O)+(-2,-0.18)$)
{$\beta = \vec u $};


\begin{scope}[shift={(10.35,0)},scale=0.88]

    \fill[white]
    (-0.34,0)
    .. controls (-0.15,0.22) and (0.18,0.22)
    .. (0.38,0)
    .. controls (0.18,-0.20) and (-0.15,-0.20)
    .. (-0.34,0)
    -- cycle;

    \fill[observerplane,opacity=0.45]
    (-0.055,0) circle (0.125);

    \draw[black,line width=0.55pt]
    (-0.055,0) circle (0.125);

    \fill[black]
    (-0.055,0) circle (0.055);

    \fill[white]
    (-0.080,0.035) circle (0.018);

    \draw[black,line width=1.00pt,line cap=round]
    (-0.34,0)
    .. controls (-0.15,0.22) and (0.18,0.22)
    .. (0.38,0);

    \draw[black,line width=0.80pt,line cap=round]
    (-0.34,0)
    .. controls (-0.15,-0.20) and (0.18,-0.20)
    .. (0.38,0);

    \draw[black,opacity=0.48,line width=0.45pt]
    (-0.24,0.13)
    .. controls (-0.05,0.29) and (0.19,0.25)
    .. (0.31,0.09);

    \draw[black,line width=0.55pt]
    (-0.34,0)
    .. controls (-0.30,0.035) and (-0.27,0.025)
    .. (-0.245,0);

    \draw[black,line width=1.10pt,line cap=round]
    (-0.24,0.31)
    .. controls (-0.08,0.41) and (0.12,0.40)
    .. (0.29,0.30);

    \draw[black,opacity=0.35,line width=0.45pt,line cap=round]
    (-0.19,0.335)
    .. controls (-0.03,0.395) and (0.13,0.38)
    .. (0.25,0.315);

\end{scope}


\draw[dimline]
(\xS,-2.48)--(\xL,-2.48);

\draw[dimline]
(\xS,-2.37)--(\xS,-2.60);

\draw[dimline]
(\xL,-2.37)--(\xL,-2.60);

\node[fill=white,inner sep=2pt]
at ({(\xS+\xL)/2},-2.48)
{$D_{ls}$};

\draw[dimline]
(\xL,-2.48)--(\xO,-2.48);

\draw[dimline]
(\xO,-2.37)--(\xO,-2.60);

\node[fill=white,inner sep=2pt]
at ({(\xL+\xO)/2},-2.48)
{$D_l$};

\draw[dimline]
(\xS,-3.05)--(\xO,-3.05);

\draw[dimline]
(\xS,-2.94)--(\xS,-3.17);

\draw[dimline]
(\xO,-2.94)--(\xO,-3.17);

\node[fill=white,inner sep=2pt]
at ({(\xS+\xO)/2},-3.05)
{$D_s$};
\end{tikzpicture}
\caption{Schematic geometry of two gravitationally lensed neutrino trajectories propagating from the source plane to the observer through a projected lens plane containing a smooth central lens and stochastic substructure.}
\label{fig:sld_geometry}
\end{figure}
\begin{equation}
\nabla^2\phi(\vec x)
=
2\kappa(\vec x),
\end{equation}
where $\vec{x}=(x_1,x_2)$ denotes the two dimensional angular position vector on the lens plane and $\kappa$ is the projected surface density and is given in units of the critical surface density as
\begin{equation}
\kappa(\vec x)
=
\frac{\Sigma(\vec x)}{\Sigma_{\rm cr}},
\end{equation}
where the critical surface density can be given as,
\begin{equation}
\Sigma_{\rm cr}
=
\frac{c^2D_s}{4\pi G D_lD_{ls}},
\label{eq:crit_density}
\end{equation}
where $D_l$, $D_s$ and $D_{ls}$ denote the angular diameter distances between observer and lens, observer and source and lens and source, respectively. For a given source at angular position $u$ and an image at angular position $\vec x$, the time delay surface is given by,
\begin{equation}
\tau(\vec x;\vec u)
=
\frac{1+z_l}{c}
\frac{D_lD_s}{D_{ls}}
\left[
\frac{1}{2}|\vec x-\vec u|^2
-
\phi(\vec x)
\right],
\label{eq:fermat_td}
\end{equation}
where $z_l$ is the lens redshift. The first term in the square bracket represents the geometrical path length contribution whereas the second term represents the gravitational potential contribution. Images form at stationary points of the time delay surface and the stationarity condition gives the lens equation as
\begin{equation}
\vec u = \vec x - \alpha(\vec x),
\qquad
\alpha(\vec x) = \nabla\phi(\vec x).
\label{eq:lens_equation}
\end{equation}
Although the time delay formalism is derived for null rays, it can be applied to ultra relativistic neutrinos at leading order. Since for $m_i^2/E^2\ll1$, the difference between the neutrino trajectory and the corresponding null trajectory is subleading. We therefore separate the lensing  induced delay which is controlled by the geometry and gravitational potential from the kinematic delay associated with the finite neutrino mass (see Ref.~\citep{Jia:2019hih}). The former determines the gravitationally induced path and potential delay whereas the latter contributes to the usual mass dependent oscillation phase. For neutrino flavours to oscillate, the phase difference between propagating mass eigenstates has to be non-zero. The perturbation in the lensing time delay induces a corresponding perturbation in the relative oscillation phase which is given as following
\begin{equation}
\delta\Phi_{ij}^{\rm lens}
=
\frac{\Delta m_{ij}^2}{2E}\,
\delta\tau_{\rm lens},
\label{eq:phase_td_relation}
\end{equation}
Thus, any stochastic fluctuation in the lensing time delay appears as a stochastic fluctuation in the oscillation phase. In realistic lensing systems, the gravitational potential is not perfectly smooth. Following the stochastic substructure description, we decompose the total lensing potential into a smooth component and a substructure component~\cite{Keeton:2009ua, DiazRivero:2017xkd, DiazRivero:2018oxk} as follows
\begin{equation}
\phi(\vec x)
=
\phi_{\rm smth}(\vec x)
+
\phi_{\rm sub}(\vec x).
\label{eq:smooth_sub_split}
\end{equation}
The smooth component $\phi_{\rm smth}$ stands for the dominant macroscopic lens potential whereas $\phi_{\rm sub}$ is generated by unresolved compact clumps. The detailed positions and masses of these clumps are not known event by event and are therefore treated statistically. Consequently, $\phi_{\rm sub}$ induces stochastic perturbations in the lensing time delay and hence in the neutrino oscillation phase. For two lensed images at angular positions $\vec x_A$ and $\vec x_B$, the difference in substructure induced time delay is defined as
\begin{equation}
\Delta\tau_{\rm sub}
=
\tau_{\rm sub}(\vec x_A)
-
\tau_{\rm sub}(\vec x_B).
\label{eq:td_diff}
\end{equation}
Since, the image positions are stationary points of the Fermat surface, the leading correction due to a small perturbing potential can be evaluated at the unperturbed image positions. Now, the corrections arising from the corresponding shifts of the image positions enter at higher order. Therefore, to leading order in the stochastic substructure potential, we can readily write
\begin{equation}
\Delta\tau_{\rm sub}
=
-
\frac{1+z_l}{c}
\frac{D_lD_s}{D_{ls}}
\Delta\phi_{\rm sub},
\label{eq:td_sub}
\end{equation}
where
\begin{equation}
\Delta\phi_{\rm sub}
=
\phi_{\rm sub}(\vec x_A)
-
\phi_{\rm sub}(\vec x_B).
\label{eq:pot_td_sub}
\end{equation}
The overall sign in Eq.~\eqref{eq:td_sub} has no physical effect on the decoherence calculation, since the relevant quantity is the variance of the stochastic time delay difference. At this point, it is important to emphasize that the observable quantity is the potential difference $\Delta\phi_{\rm sub}$ rather than the absolute value of the perturbing potential. This removes any ambiguity associated with the arbitrary zero point of the lensing potential. Now, for a point mass clump at position $\vec w_i$, the substructure potential can be written as~\citep{Keeton:2009ua}
\begin{equation}
\phi_i(\vec w)
=
\frac{m_i}{\pi}
\ln\frac{|\vec w-\vec w_i|}{a},
\label{eq:point_mass_pot}
\end{equation}
where $m_i=M_i/\Sigma_{\rm cr}$ is the clump mass scaled by the critical surface density and $a$ is an arbitrary reference scale fixing the zero of the potential. The difference between two image positions now can be given as
\begin{equation}
\phi_i(\vec w_A)-\phi_i(\vec w_B)
=
\frac{m_i}{\pi}
\ln
\left(
\frac{|\vec w_A-\vec w_i|}
{|\vec w_B-\vec w_i|}
\right),
\label{eq:point_mass_pot_diff}
\end{equation}
so the arbitrary scale $a$ cancels out. The corresponding time delay perturbation is therefore independent of the potential zero point. The variance of the stochastic time delay difference is then related to the variance of the substructure potential difference as
\begin{equation}
{\rm Var}\!\left[\Delta\tau_{\rm sub}\right]
=
\left[
\frac{1+z_l}{c}
\frac{D_lD_s}{D_{ls}}
\right]^2
{\rm Var}\!\left[\Delta\phi_{\rm sub}\right].
\label{eq:td_var}
\end{equation}
Using Eq.~\eqref{eq:phase_td_relation}, the corresponding variance of the lensing induced neutrino phase fluctuation becomes
\begin{equation}
{\rm Var}\!\left[\delta\Phi_{ij}^{\rm lens}\right]
=
\left(
\frac{\Delta m_{ij}^2}{2E}
\right)^2
{\rm Var}\!\left[\Delta\tau_{\rm sub}\right].
\label{eq:phase_var_td}
\end{equation}
Equations~\eqref{eq:td_var} and~\eqref{eq:phase_var_td} serve as the bridge between stochastic gravitational lensing and neutrino flavour decoherence. The unresolved substructure does not only shift the coherent lensing phase but also it introduces a statistical spread in the phase. Similar statistical averaging ideas appear in studies of fluctuating lightcones, where metric fluctuations smear the sharp propagation structure and the variance of the geometrical fluctuation controls the width of the resulting distribution~\cite{Ford:1994cr, Ford:1996qc, Ford:1997zb, Hu:1999mm, Yu:1999pq}. In the present work, the source of the fluctuation is not quantum space-time foam, but classical unresolved lens substructure. The decoherence effect therefore arises as an effective ensemble averaged loss of phase coherence induced by stochastic gravitational lensing.

\subsection{Lensing Induced Decoherence from Stochastic Substructure}
\label{sec:lensing_decoherence}

In this section, we derive the effective decoherence strength parameter arising from averaging the stochastic fluctuations in the neutrino phase because of the substructure presentin line of sight of neutrinos. Having established the relation between the substructure induced time delay variation and the neutrino oscillation phase, we now formulate the corresponding lensing induced decoherence effect following the prescription outlined in Ref.~\cite{Ford:1994cr, Ford:1996qc, Ford:1997zb, Hu:1999mm}. Using Eq.~\eqref{eq:phase_td_relation} and Eq.~\eqref{eq:td_sub}, the lensing induced fluctuation in the phase can be written as
\begin{equation}
\delta\Phi_{ij}^{\rm lens}
=
\frac{\Delta m^2_{ij}}{2E}
\Delta\tau_{\rm sub},
\label{eq:dphi_lens}
\end{equation}
Thus, it can be seen that the lensing contribution is geometrical and is controlled by the stochastic time delay perturbation generated by intervening substructure mass. Now, the variance of the substructure potential difference between two image positions $\vec w_A$ and $\vec w_B$ is given as~\citep{Keeton:2009ua}
\begin{align}
\mathrm{Var}\!\left[\Delta\phi_{\rm sub}\right]
&=
\mathrm{Var}\!\left[\phi_{\rm sub}(\vec w_A)\right]
+
\mathrm{Var}\!\left[\phi_{\rm sub}(\vec w_B)\right]
\nonumber\\
&\hspace{1cm}
-
2\,\mathrm{Cov}
\left[
\phi_{\rm sub}(\vec w_A),
\phi_{\rm sub}(\vec w_B)
\right],
\label{eq:dphi_var2}
\end{align}
where, the difference in substructure potential is given as
\begin{equation}
\Delta\phi_{\rm sub}
=
\phi_{\rm sub}(\vec w_A)
-
\phi_{\rm sub}(\vec w_B).
\end{equation}
We now adopt the long range and weak perturbation approximation for the clump population. In this regime, the clumps are projected at characteristic distances \(r\gtrsim R_0\) from the images and the image separation satisfies the relation
\begin{equation}
|\vec w_A-\vec w_B|
\ll
R_0.
\end{equation}
Now, expanding the substructure potential tidally up to the leading term in long range contribution in the image separation, we can obtain 
\begin{equation}
\mathrm{Var}\!\left[\Delta\phi_{\rm sub}\right]
\simeq
\frac{m_{\rm eff}}{\pi}
K_2
|\vec w_A-\vec w_B|^2,
\label{eq:dphi_var_tidal}
\end{equation}
where, the effective scaled clump mass and the convergence is given as
\begin{equation}
m_{\rm eff}
=
\frac{\langle m^2\rangle}{\langle m\rangle},
\qquad
K_2
=
\int_{R_0}^{R_{\rm max}}
\kappa_{\rm sub}(r)\,
\frac{dr}{r}.
\label{eq:meff_K2_def}
\end{equation}
Here, $m_{\rm eff}$ is the effective clump mass that controls the strength of the stochastic lensing perturbation whereas $K_2$ encodes the projected substructure convergence in the relevant annular region. Now, using Eqs.~\eqref{eq:td_var} and~\eqref{eq:phase_var_td}, the variance of the substructure induced phase difference is given as
\begin{equation}
\mathrm{Var}\!\left[\delta\Phi_{ij}^{\rm lens}\right]
=
\left(
\frac{\Delta m^2_{ij}}{2E}
\right)^2
\left[
\frac{1+z_l}{c}
\frac{D_lD_s}{D_{ls}}
\right]^2
\frac{m_{\rm eff}}{\pi}
K_2
|\vec w_A-\vec w_B|^2.
\label{eq:lens_phase_variance_tidal}
\end{equation}
To identify the stochastic substructure induced neutrino phase smearing as an effective decoherence channel, we can readily define
\begin{equation}
\mathrm{Var}\!\left[\delta\Phi_{ij}^{\rm lens}\right]
=
\Gamma_{ij}^{\rm lens}L.
\label{eq:dissipator}
\end{equation}
This gives the general expression for the decoherence strength parameter as
\begin{equation}
\Gamma_{ij}^{\rm lens}
=
\frac{1}{L}
\left(
\frac{\Delta m^2_{ij}}{2E}
\right)^2
\left[
\frac{1+z_l}{c}
\frac{D_lD_s}{D_{ls}}
\right]^2
\frac{m_{\rm eff}}{\pi}
K_2
|\vec w_A-\vec w_B|^2.
\label{eq:dissipator2}
\end{equation}
With this convention, $\Gamma_{ij}^{\rm lens}$ measures the phase variance per unit propagation length. Since Eq.~\eqref{eq:dissipator2} has a singularity at $L = 0$, we introduce a turn on scale to avoid this. With this function, we have our dissipator of the form
\begin{equation}
\Gamma_{ij}^{\rm lens}
=
\frac{1}{L}
\left[1-\exp\left(-\frac{L^2}{L_{\rm lens}^2}\right)\right]
\left(
\frac{\Delta m^2_{ij}}{2E}
\right)^2
\left[
\frac{1+z_l}{c}
\frac{D_lD_s}{D_{ls}}
\right]^2
\frac{m_{\rm eff}}{\pi}
K_2
|\vec w_A-\vec w_B|^2,
\label{eq:dissipator}
\end{equation}
where $ L_{\rm lens} $ characterizes the spatial length over which the stochastic lensing contribution becomes effective. Since, the gravitational perturbations associated with the lens and its substructure are localized along the neutrino line of sight, the corresponding dissipative contribution should act only where the neutrino traverses the lensing environment. Prior to entering this region and after escaping this region, the neutrino undergoes ordinary vacuum propagation. At this point, we emphasize that, this suppression is an effective decoherence effect as it reflects a loss of phase coherence at the ensemble level though it is not a fundamental non-unitary evolution of a neutrino state. Furthermore, in this study, $m_{\rm eff}$ characterizes the effective mass scale of the population responsible for fluctuations in the lensing potential. Following the convention adopted in Ref.~\cite{Keeton:2009ua}, the lensing mass is expressed in rescaled units as $m = M / \Sigma_{\rm cr}$ with $\Sigma_{\rm cr}$ is the critical surface mass density given in Eq.~\eqref{eq:crit_density}. The effective rescaled mass is defined by Eq.~\eqref{eq:meff_K2_def} which corresponds to effective physical mass as given by
\begin{equation}
M_{\rm eff}
=
\frac{\langle M^2\rangle}{\langle M\rangle}
=
\Sigma_{\rm cr}\,m_{\rm eff}.
\label{eq:effective_physical_mass}
\end{equation}
In this work, we often use it while presenting our results in terms of solar mass unit i.e., $M_{\rm eff} / M_\odot.$ Now, to have an astrophysical context for the explored range in this study, $ 10^{9}M_\odot \lesssim M_{\rm eff} \lesssim 10^{17}M_\odot, $ it is useful to compare these effective masses with the conventional halo mass $M_{200}$. It represents the mass enclosed within a virial radius $R_{200}$ whose mean density is $200$ times the critical density of the Universe discussed in Refs.~\cite{Navarro:1996gj, Chen:2025jch, Tiruvaskar:2025lkq, Alonso-Alvarez:2024gdz},
\begin{equation}
M_{200}
=
\frac{4\pi}{3}R_{200}^{3}
\left[200\,\rho_{\rm c}(z)\right],
\qquad
\rho_{\rm c}(z)
=
\frac{3H^2(z)}{8\pi G},
\end{equation}
with $\rho_{\rm c}$ is the critical density, $H(z)$ is the Hubble expansion rate at that epoch and $G$ is Newton's gravitational constant. Equivalently, we can write
\begin{equation}
R_{200}
=
\left[
\frac{G M_{200}}{100H^2(z)}
\right]^{1/3}.
\label{eq:R200M200}
\end{equation}
From Eq.~\eqref{eq:R200M200}, we can estimate the characteristic spatial scales associated with the effective mass range considered in our analysis by comparing $M_{\rm eff}$ with an equivalent virial mass $M_{200}$. Here, masses in the range $10^{9}M_\odot\lesssim M_{\rm eff}\lesssim10^{13}M_\odot$ corresponds to a few hundred kiloparsecs representing galactic scale environments. For $10^{14}M_\odot\lesssim M_{\rm eff}\lesssim10^{15}M_\odot$, the corresponding radii are of order a megaparsec linked to massive groups and galaxy clusters. At the upper end, $10^{16}M_\odot\lesssim M_{\rm eff}\lesssim10^{17}M_\odot$ maps to scales of several megaparsecs which is indicative of highly extended large-scale environments. However, we emphasize that this comparison does not imply the identification $M_{\rm eff}=M_{200}$ rather it is introduced only to provide an astrophysical relatibility of scale. This discussion is consistent with the hierarchical structure of massive dark matter environments discussed in Ref.~\cite{Alonso-Alvarez:2024gdz} in which
the central compact object, the inner dark matter enhancement and the
extended host halo are characterized by distinct mass and spatial
scales.

\section{Results and Discussion}
\label{sec:resanddisc}
In this section, we analyze the observables and parameters related to stochastic lensing induced decoherence. We discuss how these stochastic fluctuations affect oscillation probabilities in Earth based detectors and their directional dependence which may lead to observable variations in flavour transitions.

\subsection{Observables and Relevant Parameters}
\label{subsec:obsandpara}
The list of observables emanating from this study, which are of interest as well as the relevant parameters that affect them are listed as follows.
\begin{enumerate}
\item \textbf{Flavour Compositions:} A key observable for investigating stochastic lensing induced decoherence is the flavour composition at neutrino telescopes. This can shed light into neutrino sources and the effects of decoherence further revealing how foreground and dark matter (sub) structures impact neutrino propagation.

 \item \textbf{Oscillation Probabilities:} Studying neutrino survival and transition probabilities can shed light on decoherence strength parameters, flavour equilibration and quantify the distortions in oscillations caused by decoherence framework.
 
 \item \textbf{Flavour Displacement :} It tells us how much does the stochastic lensing move the observed flavour point in the ternary. If the shift is zero, then stochastic lensing has no observable flavour effect, otherwise it has.

\end{enumerate}

\subsection{Astrophysical Implications}
\label{subsec:astroimp}

In this section, we select two neutrino associated blazars PKS~1424+240 and TXS~0506+056 as benchmark sources and discuss whether foreground gravitational structure along their lines of sight can induce observable stochastic lensing decoherence. The lensing framework developed in this work is formulated independently of a particular astrophysical source. Since, its potential observational relevance is connected to distant high energy neutrino (HENs) sources, for which the baseline and the intervening gravitational environment gives us the condition under which stochastic phase perturbations may accumulate. Two particularly interesting sources in this context are TXS~0506+056~\citep{IceCube:2018cha, IceCube:2018dnn, Yang:2024bsf, Fiorillo:2025cgm, VERITAS:2025lvj, Stathopoulos:2026hnf} and PKS~1424+240~\citep{VERITAS:2009lpb, IceCube:2021slf, Sahu:2024ata, Padovani:2022wjk, Rovero:2016igo, Rovero:2015ksa}. At this point, we emphasize that, this discussion is intended only to illustrate the astrophysical regime in which the mechanism studied here may become relevant where a source-by-source reconstruction of the foreground mass distribution, statistics and the resulting lensing induced flavour modification is beyond the scope and interest of the present work. Here, TXS~0506+056 becomes important as it constitutes the most widely studied blazar associated with HENs emission. The source is located at $z\simeq0.3365$ and was spatially as well as temporally associated with IceCube~170922A~\citep{IceCube:2018cha, IceCube:2018dnn} whereas an earlier excess of neutrino events was identified in existing IceCube data~\citep{IceCube:2018cha}. Its multi-messenger behaviour has motivated a broad range neutrino production models with TeV to PeV ranges of energy. Hence,for the present work, it provides an observationally established example of a cosmological HEN source for which propagation induced effects on flavour coherence can in principle be testified or falsified.

Similarly, PKS~1424+240 serves as a complementary example. Spectroscopic observations place the source at $z\simeq0.6047$ making it substantially more distant than TXS~0506+056 with comparable neutrino energies~\cite{Padovani:2022wjk}. Additionally, studies of its environment have identified a galaxy system at $z\simeq0.60$ associated with the blazar and have revealed intervening galaxies along the same line of sight including a possible foreground association near $z\simeq0.47$.~\cite{Rovero:2016igo, Rovero:2015ksa}. Hence, this becomes interesting for the current framework as the lensing induced phase variance depends not only on the distance from where the lensing effect begins but also on the projected gravitational structure encountered during propagation. PKS~1424+240 therefore presents a useful counterpart to TXS~0506+056 for illustrating how different source distances and foreground environments may lead to different levels of stochastic phase alteration resulting in decoherence. In a nut shell, two neutrino sources TXS~0506+056 and PKS~1424+240 are located at different redshifts hence probe different foreground gravitational environments. The comparison between the two can therefore be interesting where the former provides an observationally established neutrino source benchmark and the latter offers a longer propagation baseline and a line of sight with identified galaxy structure. We however stress that the present analysis does not predict a specific measurement for either TXS~0506+056 or PKS~1424+240. At this point, we emphasize that the quoted source redshifts determine the full cosmological source to observer geometry whereas the propagation length $L$ used in our analysis denotes only the local interval over which the stochastic lensing effect becomes active. Outside this region, the neutrino is assumed to undergo ordinary vacuum propagation. Moreover, the results presented in the following sections are therefore discussed in the context of the sources mentioned above, which serve as representative astrophysical systems for analyzing the effects of lensing induced neutrino decoherence.

\subsection{Flavour Composition at Neutrino Telescopes}
\label{subsec:flavcomp}
As discussed in earlier sections, neutrinos travelling long distances before reaching Earth serve as best bet messengers for illustrating decoherence effects. Additionally, different sources have been modelled to produce various neutrino compositions. Hence, we stress on decoherence effects on final flux composition of the neutrinos coming from various sources of astrophysical origin~\cite{Anchordoqui:2005is, Beacom:2003nh, Choubey:2009jq}. We assume these sources will generate fluxes of electron, muon and tau neutrinos shown as  $f_e^{(0)}$, $f_\mu^{(0)}$ and $f_\tau^{(0)}$ respectively. Due to oscillations and eventually decoherence the initial flux composition no longer remains the same as neutrinos arrive at the observatories on Earth.
\begin{figure}[t]
\centering
\begin{subfigure}{0.49\linewidth}
\centering
\includegraphics[width=\linewidth]{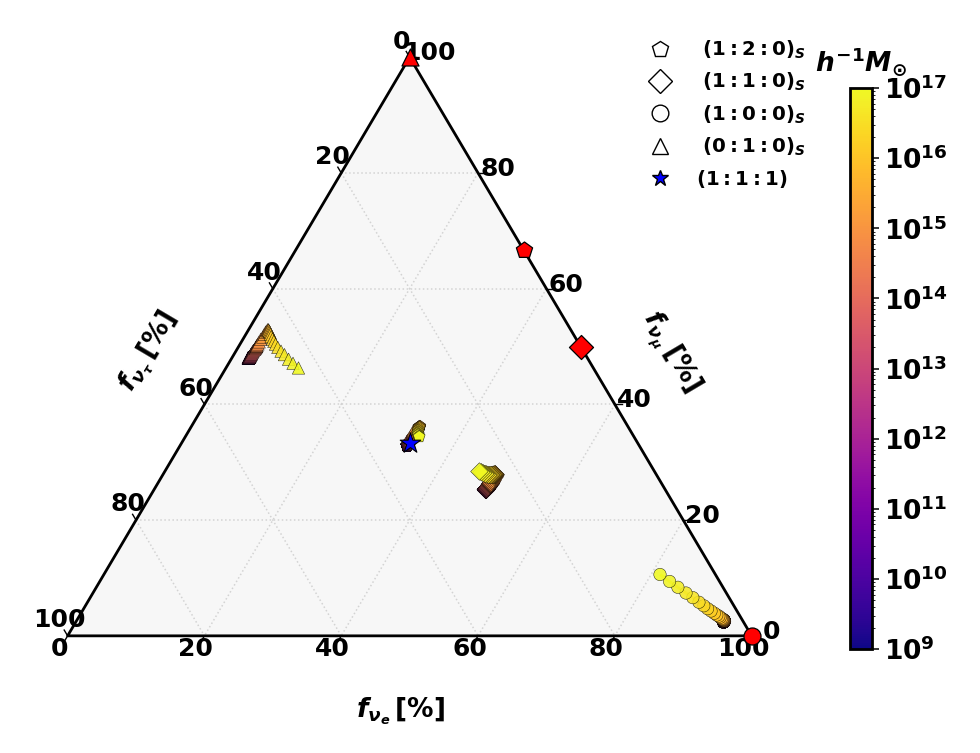}
\caption{$E_{\nu}$=10 PeV}
\label{fig:meff10PeV}
\end{subfigure}
\hfill
\begin{subfigure}{0.49\linewidth}
\centering
\includegraphics[width=\linewidth]{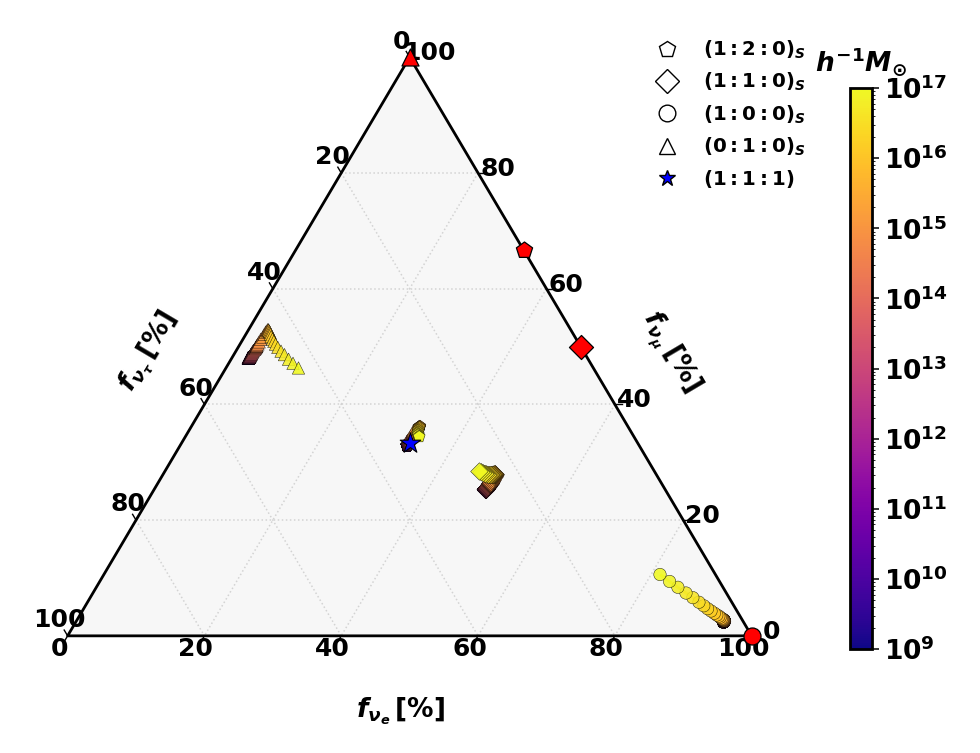}
\caption{$E_{\nu}$=100 PeV}\label{fig:meff100PeV}
\end{subfigure}
\caption{Ternary representation of the flavour composition at Earth for $E_\nu=10~{\rm PeV}$ and $E_\nu=100~{\rm PeV}$ obtained by varying the effective lensing mass $M_{\rm eff}$. The colour bar denotes $M_{\rm eff}$ in units of $h^{-1}M_\odot$. The different markers correspond to different initial source compositions, whereas the blue star represents exact flavour equipartition $(1:1:1)$.}
\label{fig:meff_variation}
\end{figure}
In Fig.~\ref{fig:meff10PeV}, we present the predicted flavour composition of neutrinos arriving at Earth with a fixed energy of $E_\nu = 10~{\rm PeV}$ while varying the effective lensing mass scale $M_{\rm eff}$. The ternary diagram illustrates the distribution of the three flavour fractions $(f_{\nu_e}, f_{\nu_\mu}$ and $ f_{\nu_\tau})$ constrained by the condition $f_{\nu_e} + f_{\nu_\mu} + f_{\nu_\tau} = 1$. We showcase four distinct trajectories that represent different possible source compositions as $(1:2:0)$, $(1:1:0)$, $(1:0:0)$ and $(0:1:0)$. These trajectories correspond to initial flavour compositions associated with pion beams, charm production, neutron beams and muon-damped sources respectively. Notably, the blue star indicates the democratic flavour composition $(f_{\nu_e}:f_{\nu_\mu}:f_{\nu_\tau}) = (1:1:1)$ which serves as a standard reference point for complete flavour equipartition. The colour gradient represents the variation in the effective lensing mass scale ranging from lower to higher values with the colour bar indicating the effective mass variation i.e., $M_{\rm eff}\sim10^9-10^{17}h^{-1}M_\odot$. As $M_{\rm eff}$ increases, the stochastic phase dispersion induced by lensing becomes more pronounced. This results in the predicted flavour composition of neutrinos to move along source dependent trajectories in the ternary plane. The separation and direction of these trajectories provide a quantifiable measure of how each source composition responds to the decoherence effects caused by lensing. For pion beam source represented as $(1:2:0)$, the flavour composition remains close to the equipartition point. At lower values of $M_{\rm eff}$, the composition is approximately $(0.33:0.32:0.34)$ closely mirroring with $(1:1:1)$. As $M_{\rm eff}$ rises, the trajectory shifts slightly towards a larger fraction of electron and muon flavours whereas the tau fraction decreases which results in a composition of roughly $(0.34:0.36:0.30)$ at the high end of the mass range. This indicates that for pion beam sources, stochastic lensing decoherence has a mild effect at an energy of $E_\nu=10~{\rm PeV}$ with the compositions remaining near flavour equipartition. On the other hand, the charm like source composition $(1:1:0)$ starts at a point away from equipartition around $(0.49:0.25:0.26)$ and shifts towards $(0.45:0.29:0.26)$ as $M_{\rm eff}$ increases. Here, the major change involves a reduction in the electron flavour fraction and an enhancement of the muon flavour component wheras the tau flavour fraction remains comparatively stable. This indicates that the $(1:1:0)$ source composition is moderately sensitive to variations in the effective lensing mass scale. The neutron beam source $(1:0:0)$ displays one of the most significant responses. Initially, at the lower end of the $M_{\rm eff}$ range, the flavour composition is highly dominated by electrons i.e. $(0.95:0.02:0.03)$. As $M_{\rm eff}$ increases, there is a significant decrease in the electron flavour fraction with corresponding increases in the muon and tau fractions. At the highest value for $M_{\rm eff}$, the composition shifts to about $(0.80:0.11:0.09)$. Thus, this source type is highly sensitive to lensing induced changes in flavour compositions. Finally, the muon-damped like source $(0:1:0)$ also showcases a significant shift in the ternary plane. For smaller $M_{\rm eff}$, the composition is concentrated near the $\nu_\mu-\nu_\tau$ side of the triangle around $(0.02:0.48:0.50)$. As the effective lensing mass increases, the electron flavour fraction rises whereas the tau flavour fraction goes down leading to a flavour composition of approximately $(0.11:0.45:0.44)$ at the upper end of the mass range. This behaviour illustrates that sources with initially suppressed electron flavour can gain a substantial electron component during propagation in light of stochastic lensing induced decoherence. In a nutshell, Fig.~\ref{fig:meff10PeV} shows that the lensing induced decoherence effect is source dependent. The pion beam source remains close to flavour equipartition and is therefore less promising in this benchmark. In contrast, the neutron beam and muon damped sources showcase much larger displacements in the ternary plane as $M_{\rm eff}$ is varied. This suggests that different source compositions provides us a cleaner probe of stochastic lensing induced modifications to astrophysical neutrino flavour ratios. However, We here note that, the figure should be interpreted as a benchmark level theoretical result at fixed $E_\nu=10~{\rm PeV}$. In this work, we are not considering the flavour reconstruction uncertainties and event statistics. In comparision to the $10~{\rm PeV}$ figure, in Fig.~\ref{fig:meff100PeV}, we show the $100~{\rm PeV}$ case as it probes astrophysical energy regime relevant for upcoming high energy neutrino telescopes. The qualitative behaviour remains similar as the pion beam source stays close to flavour equipartition whereas the neutron beam and muon damped sources show larger displacements in the ternary plane. However, the importance of the $100~{\rm PeV}$ case is that any visible displacement at such high energy would indicate that the stochastic lensing contribution remains effective even in the ultra high energy regime. Since, the lensing induced phase fluctuation contains the factor $(\Delta m^2_{ij}/2E)^2$, the decoherence effect is generally expected to be weaker at $100~{\rm PeV}$ than at $1~{\rm TeV}$ for the same lensing configuration. Therefore, a sizeable shift in the $100~{\rm PeV}$ flavour composition points towards stronger effective lensing environments, larger $M_{\rm eff}$ or likely sufficiently long propagation distances. Hence, it can be seen that even $100~{\rm PeV}$ neutrinos serve as a promising experimental probe to the lensing induced decoherence.

\begin{figure}[t]
\centering
\begin{subfigure}{0.49\linewidth}
\centering
\includegraphics[width=\linewidth]{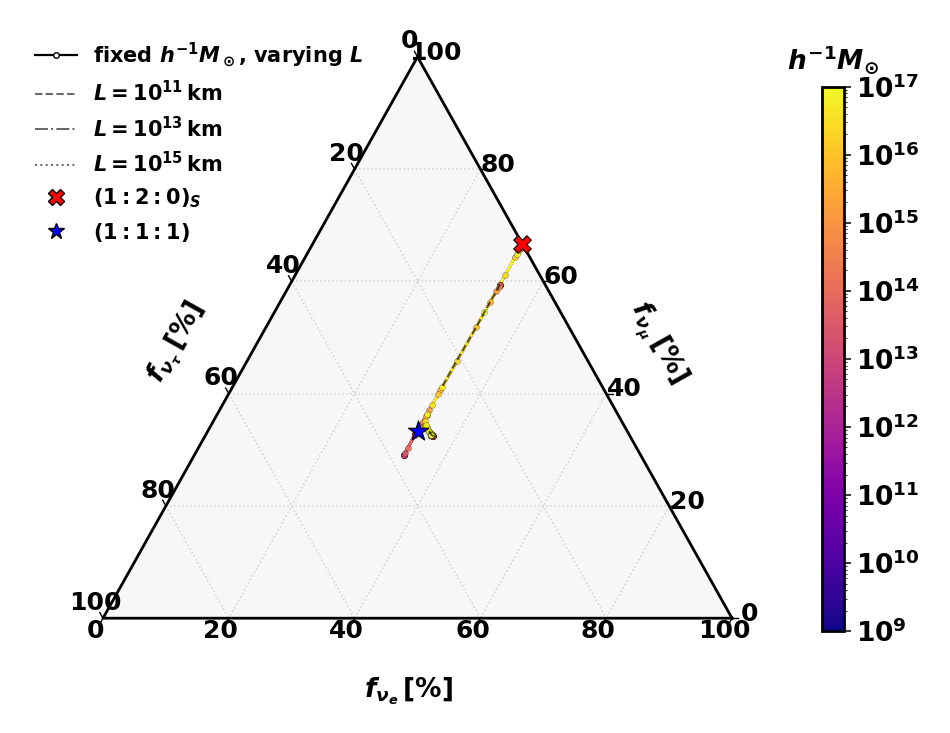}
\caption{Source $(1: 2: 0)$}
\label{fig:lmeff_120_10PeV}
\end{subfigure}
\hfill
\begin{subfigure}{0.49\linewidth}
\centering
\includegraphics[width=\linewidth]{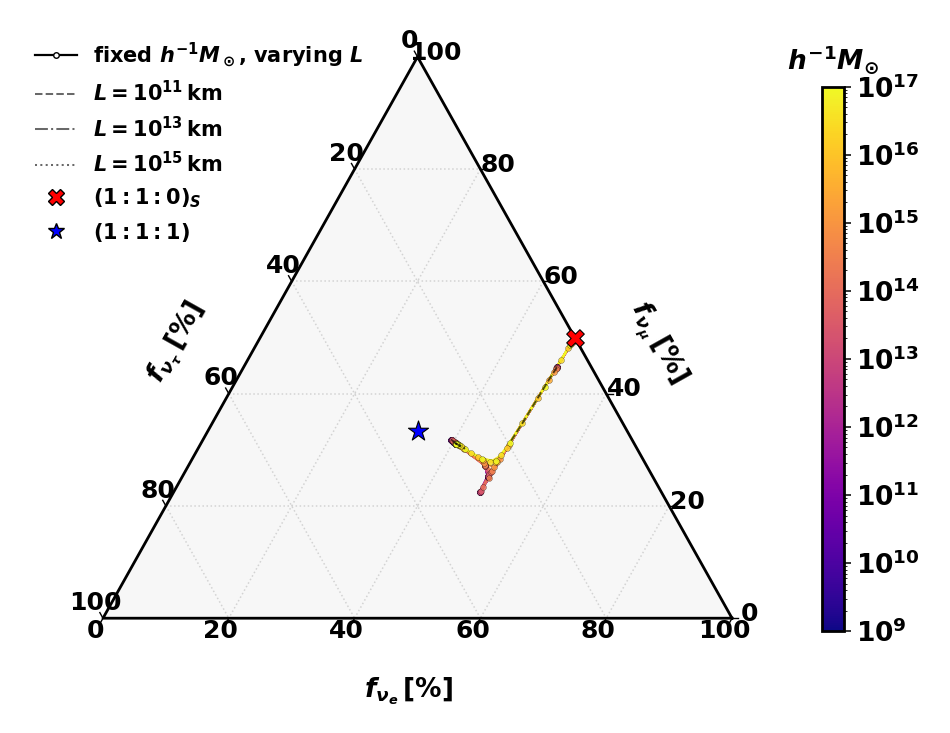}
\caption{Source $(1: 1: 0)$}
\label{fig:lmeff_110_10PeV}
\end{subfigure}
\vspace{0.25cm}

\begin{subfigure}{0.49\linewidth}
\centering
\includegraphics[width=\linewidth]{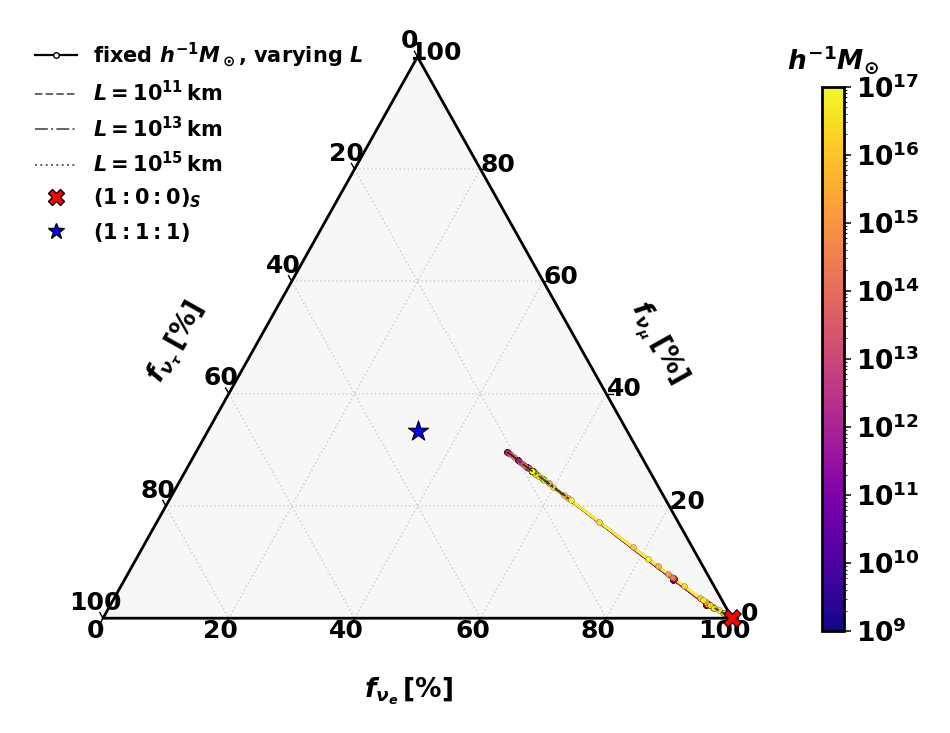}
\caption{Source $(1: 0: 0)$}
\label{fig:lmeff_100_10PeV}
\end{subfigure}
\hfill
\begin{subfigure}{0.49\linewidth}
\centering
\includegraphics[width=\linewidth]{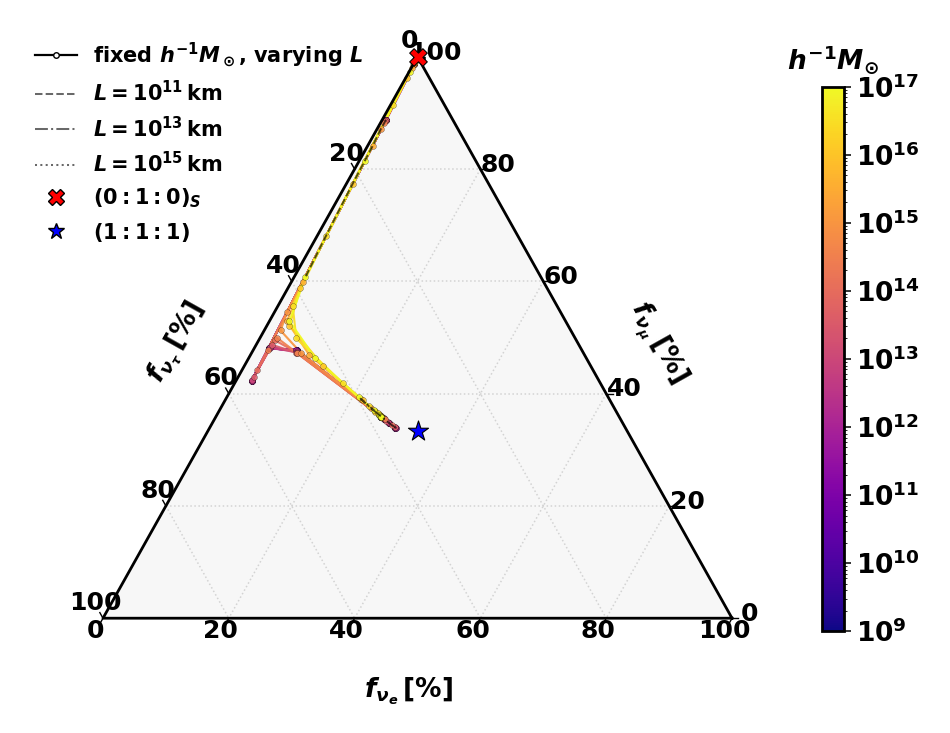}
\caption{Source $(0: 1: 0)$}
\label{fig:lmeff_010_10PeV}
\end{subfigure}
\caption{Ternary representation of the flavour composition at Earth for different initial source compositions at $E_\nu=10~{\rm PeV}$. The effective lensing mass $M_{\rm eff}$ is varied over the range $M_{\rm eff}\sim10^9-10^{17}h^{-1}M_\odot$ whereas selected propagation lengths $L=10^{11}~{\rm km}$, $10^{13}~{\rm km}$ and $10^{15}~{\rm km}$ are shown using different line styles. The red cross denotes the corresponding source composition whereas the blue star denotes flavour equipartition $(1:1:1)$.}
\label{fig:lmeff10PeV}
\end{figure}

Moreover, in Fig.~\ref{fig:lmeff10PeV}, we showcase the combined dependence of the flavour composition on the effective lensing mass $M_{\rm eff}$ and the propagation length $L$ for four different source compositions. The data files used for these plots contain five columns, corresponding to the three flavour fractions, $M_{\rm eff}$ and the selected baseline $L$. In this figure, instead of displaying the full continuous baseline scan, we choose three representative propagation lengths, $L=10^{11}~{\rm km}$, $10^{13}~{\rm km}$ and $10^{15}~{\rm km}$ and show how the flavour ratios move in the ternary plane as $M_{\rm eff}$ is changed. The colour gradient represents the variation of $M_{\rm eff}$ from $10^9h^{-1}M_\odot$ to $10^{17}h^{-1}M_\odot$. For a fixed value of $L$, increasing $M_{\rm eff}$ enhances the stochastic lensing induced phase perturbations and shifts the predicted flavour composition away from its source point. On the other hand, comparing different line styles at a fixed $M_{\rm eff}$ signifies the role of propagation length. Larger $L$ allows the lensing induced phase fluctuations to accumulate over a longer path and hence produces a stronger shift in the ternary plane. For pion beam source $(1:2:0)$ shown in Fig.~\ref{fig:lmeff_120_10PeV}, the trajectories move from the source point towards the equipartition indicating that this source flavour composition is relatively stable against lensing induced modifications. For charm like source $(1:1:0)$ shown in Fig.~\ref{fig:lmeff_110_10PeV}, the trajectories exhibit a wider spread around the central region. This indicates a moderate sensitivity to both $M_{\rm eff}$ and $L$ with the electron flavour fraction decreasing and the muon and tau fractions becoming more pronounced as the lensing effect increases. The neutron beam source $(1:0:0)$ shown in Fig.~\ref{fig:lmeff_100_10PeV} showcases a strong shift away from the electron flavour dominated vertex. As either $M_{\rm eff}$ or $L$ increases, the electron flavour fraction decreases and the muon and tau flavour fraction grows. This makes the neutron beam source one of the cleaner channels for visualising stochastic lensing induced flavour redistribution. The muon damped source $(0:1:0)$, shown in Fig.~\ref{fig:lmeff_010_10PeV} also shows a large displacement with the trajectory moving away from the muon flavour dominated vertex towards the interior of the ternary plane. This behaviour shows that initially flavour asymmetric sources are more sensitive to the combined variation of $M_{\rm eff}$ and $L$ than the pion beam source. Hence, in general Fig.~\ref{fig:lmeff10PeV} demonstrates that the lensing induced modification of astrophysical neutrino flavour ratios depends jointly on the effective lensing mass and the propagation length window where the interaction takes place. The role of $M_{\rm eff}$ is to control the strength of the stochastic lensing environment whereas $L$ controls the accumulation of the induced phase fluctuation during propagation. Therefore, the combined scan in $M_{\rm eff}$ and $L$ provides us a more informative benchmark than varying $M_{\rm eff}$ alone.

\subsection{Oscillation Probabilities}
\label{subsec:oscprobs}
In this section, we discuss how stochastic lensing induced decoherence impacts neutrino oscillation probabilities among different flavour states. It reduces the oscillation amplitude leading to slower and less visible transitions. These changes highlights the effects of dark matter substructure on neutrino propagation.
\begin{figure}[t]
\centering
\includegraphics[width=0.49\linewidth]{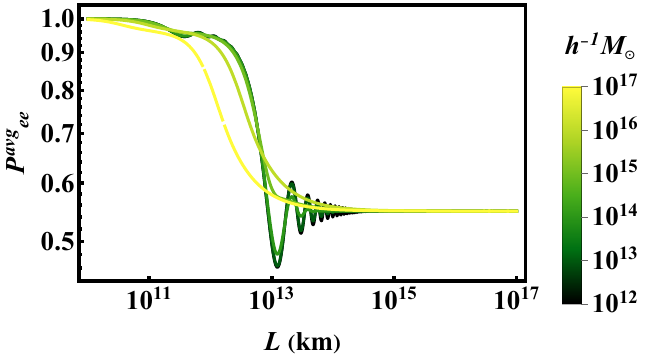}

\vspace{0.5cm}

\includegraphics[width=0.49\linewidth]{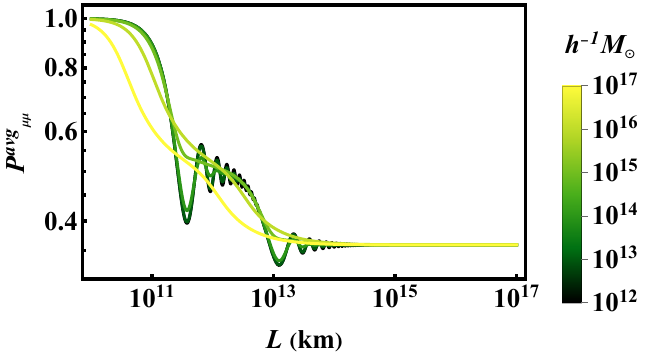}
\hfill
\includegraphics[width=0.49\linewidth]{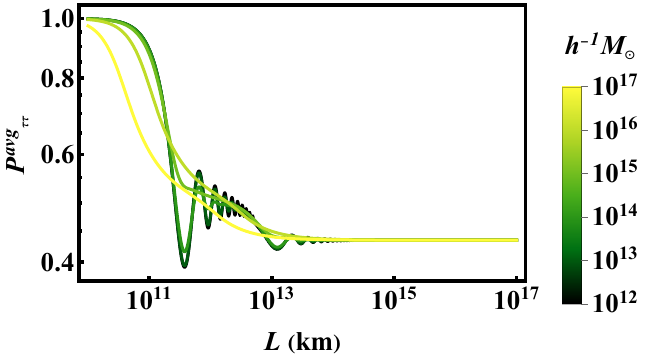}
\caption{Neutrino survival probabilities as functions of the propagation length $L$ for fixed neutrino energy $E_\nu=100~{\rm PeV}$. The upper panel shows electron neutrino survival probability whereas the lower left and lower right panels show  muon and tau neutrino survival probabilities respectively. The colour bar represents the effective lensing mass $M_{\rm eff}$ in units of $h^{-1}M_\odot$.}
\label{fig:oscprob_100PeV}
\end{figure}
It can be seen in Fig.~\ref{fig:oscprob_100PeV}, we show the averaged neutrino survival probabilities as functions of the propagation length $L$ for neutrinos with energy $E_\nu=100~{\rm PeV}$. The upper panel shows electron neutrino survival probability $P_{ee}^{\rm avg}$, whereas the lower left and lower right panels show muon and tauon neutrino survival probabilities$P_{\mu\mu}^{\rm avg}$ and $P_{\tau\tau}^{\rm avg}$ respectively. The colour gradient represents the variation of the effective lensing mass in the range $M_{\rm eff}\sim10^{12}-10^{17}h^{-1}M_\odot$. We here observe that how the stochastic lensing induced dissipator modifies the survival probabilities as the neutrino propagates over increasing baselines. In the upper panel, the electron survival probability $P_{ee}^{\rm avg}$ remains close to unity at small propagation lengths indicating that the lensing induced phase perturbations has not yet accumulated appreciably. As $L$ increases, the probability starts to decrease and eventually approaches an averaged asymptotic value around $P_{ee}^{\rm avg}\simeq0.55$. The transition occurs earlier for larger $M_{\rm eff}$, since a larger effective lensing mass enhances the stochastic time delay variance and therefore strengthens the decoherence effect. In contrast, for smaller $M_{\rm eff}$, the transition is delayed to larger baselines and residual oscillatory features remain visible before the probability settles down to its averaged value. In the lower left panel, the muon survival probability $P_{\mu\mu}^{\rm avg}$ shows a stronger suppression compared to the electron channel. Beginning from $P_{\mu\mu}^{\rm avg}\simeq1$ at small $L$, the probability decreases rapidly once the lensing induced damping becomes effective and approaches an asymptotic value around $P_{\mu\mu}^{\rm avg}\simeq0.33$. The low $M_{\rm eff}$ curves show more pronounced residual oscillatory behaviour including visible dips and recoveries because the decoherence is weaker and the oscillatory interference terms are not immediately suppressed. In contrast, for larger $M_{\rm eff}$, the damping is stronger and the transition to the averaged regime becomes smoother. In the lower right panel, the tauon survival probability $P_{\tau\tau}^{\rm avg}$ behaves similarly to the muon survival probability but approaches a slightly higher asymptotic value around $P_{\tau\tau}^{\rm avg}\simeq0.44$. At smaller baselines, $P_{\tau\tau}^{\rm avg}$ remains close to unity, whereas at larger baselines it decreases as the stochastic lensing effect accumulates. As in the other channels, larger $M_{\rm eff}$ values lead to an earlier and smoother approach to the averaged regime whereas smaller values allow residual oscillatory structures to persist over a wider range of $L$.

\subsection{Detection Prospects}
\label{subsec:detecpros}

In order to quantify the impact of lensing induced decoherence across current and future high energy neutrino observatories employing optical Cherenkov and radio detection techniques such as IceCube~\cite{IceCube:2002eys}, IceCube-Gen2~\cite{IceCube-Gen2:2020qha}, KM3NeT~\cite{KM3NeT:2024jji}, GRAND~\cite{GRAND:2018iaj}, RNO-G~\cite{RNO-G:2023kag} and POEMMA~\cite{Olinto:2023vmx} etc., we introduce a quantity that measures how far the lensing modified flavour composition is displaced from the standard averaged oscillation. In the absence of stochastic lensing, the averaged flavour composition at Earth is given by
\begin{equation}
f_\alpha^{(0)}
=
\sum_{\beta=e,\mu,\tau}
P_{\beta\alpha}^{(0)} f_\beta^S ,
\label{eq:standard_flavour_composition}
\end{equation}
where $P_{\beta\alpha}^{(0)}$ is the trivial oscillation probability. Now, in the presence of stochastic lensing induced decoherence, the flavour composition can be written as
\begin{equation}
f_\alpha^{\rm lens}
=
\sum_{\beta=e,\mu,\tau}
P_{\beta\alpha}^{\rm lens}
\left(
E_\nu,L,M_{\rm eff}
\right)
f_\beta^S .
\label{eq:lensing_flavour_composition}
\end{equation}
Here, we define the flavour displacement observable as the distance between the lensing modified flavour point and the standard oscillation flavour point as
\begin{equation}
\widehat{\Delta}_{\rm lens}
=
\frac{1}{\sqrt{2}}
\left[
\sum_{\alpha=e,\mu,\tau}
\left(
f_\alpha^{\rm lens}
-
f_\alpha^{(0)}
\right)^2
\right]^{1/2}.
\label{eq:delta_lens_normalized}
\end{equation}
where the factor $1/\sqrt{2}$ ensures normalization such that $0\leq\widehat{\Delta}_{\rm lens}\leq 1$. Hence, $\widehat{\Delta}_{\rm lens}=0$ corresponds to no observable shift from the standard averaged oscillation prediction whereas a nonzero value indicates that stochastic lensing has displaced the expected flavour composition at Earth. To construct directional dependence, we now promote the effective lensing mass to a sky dependent quantity $M_{\rm eff}(\hat n)$. We use the Planck Sunyaev-Zel'dovich (Planck-SZ) catalogue~\cite{Planck:2015koh} as a phenomenological tracer of foreground lensing environments and map the normalized SZ proxy $W_{\rm SZ}(\hat n)$ onto a directional effective mass $M_{\rm eff}(\hat n)$. The flavour displacement observable is then evaluated as $(\widehat{\Delta}_{\rm lens}(\hat n)=\widehat{\Delta}_{\rm lens}[E_\nu,L,M_{\rm eff}(\hat n)])$ for fixed neutrino energy, propagation length and source composition as
\begin{equation}
\widehat{\Delta}_{\rm lens}(\hat n)
=
\frac{1}{\sqrt{2}}
\left[
\sum_{\alpha=e,\mu,\tau}
\left(
f_\alpha^{\rm lens}
\left(E_\nu,L,M_{\rm eff}(\hat n)\right)
-
f_\alpha^{(0)}
\right)^2
\right]^{1/2}.
\label{eq:directional_delta_lens}
\end{equation}
\begin{figure}[htbp]
\centering
\includegraphics[width=1.0\linewidth]{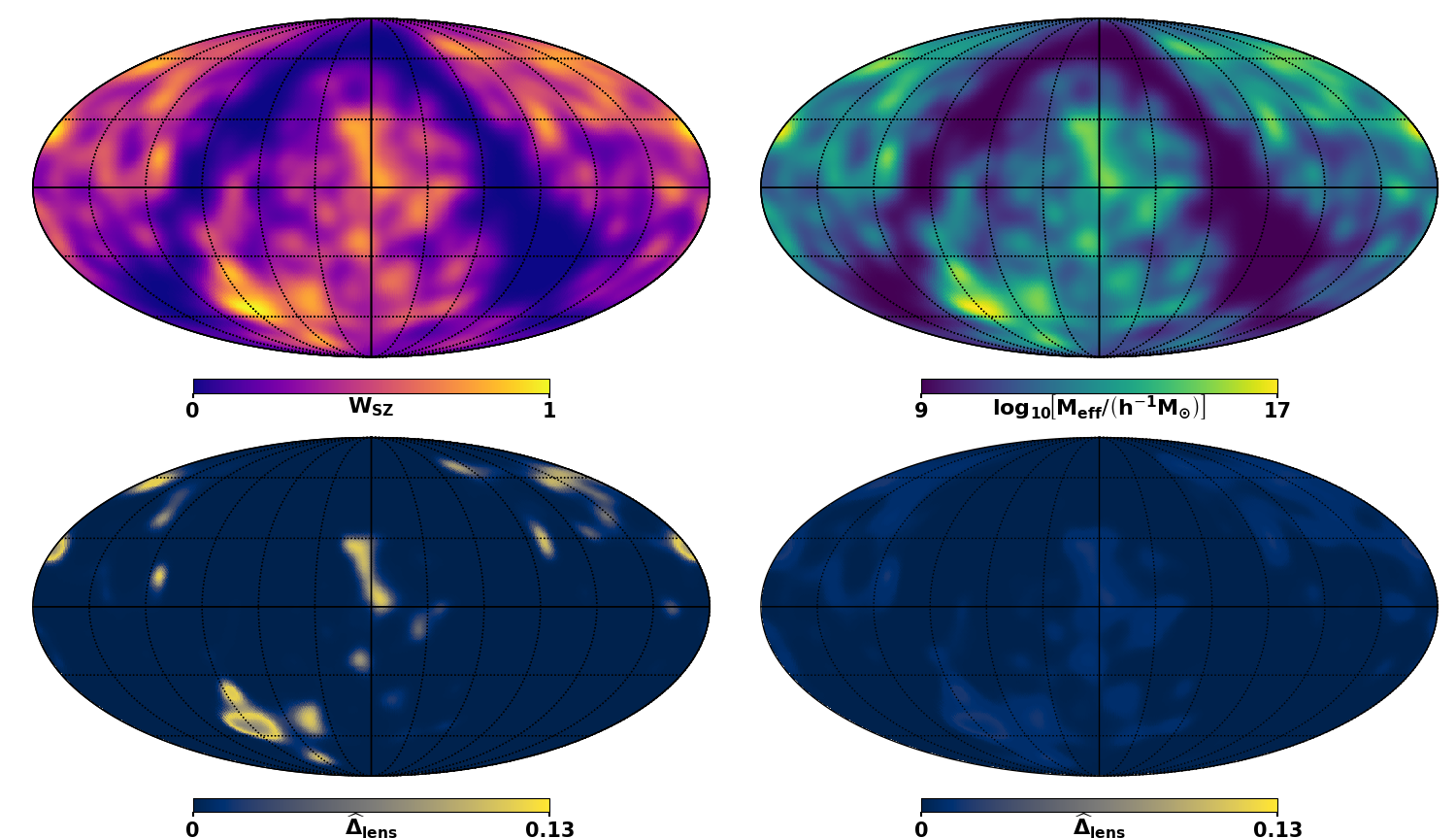}
\caption{Sky distribution of the stochastic lensing flavour displacement constructed using the Planck-SZ catalogue as a phenomenological tracer of foreground lensing environments. The top left panel shows the normalized Planck-SZ foreground structure proxy $W_{\rm SZ}(\hat n)$ whereas the top right panel shows the corresponding directional effective lensing mass displayed as $\log_{10}[M_{\rm eff}/(h^{-1}M_\odot)]$. The bottom left and bottom right panels show the directional flavour displacement  for a neutrino energy $E_\nu=100~{\rm PeV}$ with source composition $(1:0:0)$ and propagation lengths $L=10^{13}~{\rm km}$ and $L=10^{14}~{\rm km}$ respectively.}
\label{fig:skymap}
\end{figure}
In Fig.~\ref{fig:skymap}, we show the directional dependence of the stochastic lensing framework by constructing the full sky maps using the Hierarchical Equal Area isoLatitude Pixelization (HEALPix)\footnote{\url{https://healpix.sourceforge.net}}
scheme and the \texttt{healpy} package~\cite{Gorski:2004by, Zonca:2019vzt}. The top left panel shows the foreground lensing proxy constructed from the Planck-SZ cluster catalogue where the catalogue positions and mass related observables trace the large scale distribution of foreground structures. Since the Planck-SZ catalogue does not directly measure unresolved dark matter substructure, hence this map should be interpreted as a phenomenological tracer of directions in which stronger gravitational lensing environments may occur. The top right panel shows the corresponding direction dependent effective lensing mass $M_{\rm eff}(\hat n)$ which is obtained by mapping $M_{\rm eff}(\hat n)$ i.e., the foreground proxy onto the effective stochastic lensing parameter entering the neutrino decoherence framework. The bottom left and bottom right panels display the resulting flavour displacement, $\Delta_{\rm lens}(\hat n) = \Delta_{\rm lens}[E_\nu, L, M_{\rm eff}(\hat n)]$, for propagation lengths $L=10^{13}$ km and $L=10^{14}$ km respectively. This comparison illustrates how the directional flavour modification accumulates with propagation length with regions associated with larger effective lensing masses generally producing stronger deviations from the standard averaged flavour composition. As we can see, the bottom left figure shows that the lensing induced flavour displacement follows the same directional structure inherited from $M_{\rm eff}(\hat n)$  with the strongest deviations appearing along sightlines associated with larger foreground lensing weights as highlighted by the bright spots. In these regions, the flavour displacement reaches values of approximately $13\%$. The bottom right figure corresponding to the longer propagation length exhibits a generally smaller directional in contrast to the left panel. This does not imply weaker stochastic effects rather it indicates that the neutrinos are approaching the long astrophysical baseline regime from lens to observer in which further phase fluctuations produces about $4\% - 5\%$ additional change in the flavour displacement.

Furthermore, as can be inferred from Eq.~\eqref{eq:dissipator2}, the strength of the lensing induced decoherence depends on the neutrino energy and on the effective scaled substructure mass parameter $m_{\rm eff}$ together with the geometry and statistical properties of the intervening lens. Since, the interaction is localized within the spatial extent of the lensing environment, the corresponding decoherence is accumulated primarily while the neutrino traverses this region after which the lensed neutrino state propagates (almost) freely to the observer. Consequently, the magnitude of the effect is expected to depend on the particular line of sight as different arrival directions probe different foreground gravitational environments. High energy neutrino observatories such as IceCube, IceCube-Gen2, and KM3NeT may therefore provide an opportunity to search for such direction dependent modifications in neutrino phases arriving along line of sights containing sufficiently massive and structured foreground lens systems.

However, the present uncertainties remain large to identify small source-dependent shifts from the standard flavour region when uncertainties in the production flavour composition and source spectrum are included. Hence, it motivates us for a combined source-resolved and direction-dependent analysis rather than a purely diffuse flavour measurement. The directional maps constructed in this work provide a complementary way of formulating such a search. Using the Planck-SZ signal as a tracer of the foreground large scale structure, cluster rich directions of the sky associated with a larger effective lensing mass with a larger predicted flavour displacement can be used to identify priority directions whereas cluster void directions may provide an approximate control line of sight to test. Consequently, we can look for a statistical correlation between the reconstructed flavour composition and a direction dependent lensing imprint on neutrino phase rather than requiring the effect to be established from a single neutrino source. Hence, in a foreseeable future, with improving flavour, energy, and directional sensitivity in present and upcoming neutrino observatories, lensing induced decoherence could therefore be tested as a direction dependent propagation effect and potentially provide a new probe of neutrino properties through the foreground gravitational environment.

\section{Summary and Conclusion}
\label{sec:sumandconc}
In this work we have developed a phenomenological framework for neutrino decoherence induced by stochastic gravitational lensing. Rather than treating the lensing potential as fully smooth and deterministic, we decomposed it into a smooth (or coherent) background contribution and a fluctuating component associated with unresolved mass substructure. These fluctuations generate stochastic corrections to the gravitational time delay experienced by neutrinos propagating along lensed trajectories and because different mass eigenstates accumulate phase at slightly different rates, they induce random relative phases in the neutrino mixing. Assuming approximately Gaussian fluctuations and performing ensemble averaging over the unresolved substructure leads to an exponential suppression of the interference terms. The resulting behaviour may be interpreted as an effective decoherence mechanism whose strength depends on the neutrino energy, propagation distance, lens geometry, image separation, and statistical properties of the lens population. In particular, the decoherence effect scales with the squared neutrino mass splitting whereas it gets enhanced by larger effective lensing masses and becomes weaker at higher neutrino energies. Moreover, we examined the phenomenological consequences of this mechanism through flavour transition probabilities, neutrino flavour compositions, and trajectories aided with directional observables through sky maps. The survival probability analysis shows that sufficiently strong stochastic lensing can drive the flavour evolution away from the coherent oscillation regime and toward an effectively averaged limit. The ternary flavour plots likewise demonstrate that different source compositions respond differently as the effective lensing strength is varied, making the flavour triangle a useful visual tool for understanding both the magnitude and the source dependence of the effect. To quantify the observable departure from standard averaged oscillations, we introduced the flavour displacement observable $\widehat{\Delta}_{\rm lens}$, which measures the shift between the lensing modified flavour composition and the conventional expectation from trivial oscillation. It provides a compact diagnostic with which different energies, source classes, baselines, and lensing environments can be compared. For the benchmark configurations explored here, the displacement can reach a phenomenologically relevant level, suggesting that stochastic lensing may leave an observable imprint on the flavour composition of high energy astrophysical neutrinos. We also investigate the directional effect by constructing sky-dependent effective lensing maps. In this analysis, the Planck-SZ catalogue was employed as a phenomenological tracer of foreground matter rather than as a direct probe of dark matter substructure. By converting this foreground proxy into a directional effective lensing mass, we obtained corresponding maps of the flavour displacement, $\Delta_{\rm lens}(\hat n) = \Delta_{\rm lens}\left[E_\nu, L, M_{\rm eff}(\hat n)\right]$. These maps suggest that if stochastic lensing contributes appreciably to neutrino decoherence, the resulting flavour modification should not be isotropic but instead exhibit statistical correlations with foreground lensing structures along the line of sight. Cluster rich directions will show promising effects whereas cluster void directions will not. 
This work should be seen as a first phenomenological step towards quantifying realistic lensing effects in neutrino sector and it's implications. The effective lensing mass and fluctuation amplitude provide a simplified description of unresolved structures and a more realistic treatment would require explicit statistical modelling of the sub-halo mass function, spatial distribution, internal density profiles, lens-selection effects, and redshift evolution. Here, we have addressed dark matter substructure at the level of variance introduced by the stochasticity arising from perturbations in the neutrino phase. However, a detailed matter content of the substructure has not been discussed here which we leave for future work. It will also be important, in future works, to incorporate realistic neutrino source distributions, energy spectra, detector exposure, flavour reconstruction uncertainties, and finite angular resolution which we have not discussed here. Despite these limitations, the framework established here provides a direct link between stochastic gravitational lensing, neutrino phase fluctuations and observable flavour transitions. It therefore opens a complementary route for probing small scale structure in the Universe not by resolving individual sub-haloes but by searching for their cumulative statistical imprint on neutrino flavour coherence. Future analyses that combine astrophysical neutrino flavour measurements with foreground lensing catalogues may provide a novel probe of unresolved gravitational structure and upcoming observatories such as IceCube-Gen2 will be essential for testing this possibility.

\acknowledgments
BKA acknowledges the University Grants Commission (UGC), Government of India, for financial support via the UGC-NET Junior Research Fellowship. BKA thanks Santu Kumar Manna and Shamim Akhtar for fruitful discussions. Some of the results in this paper have been derived using the \texttt{healpy} and HEALPix packages. UKD acknowledge support from the Anusandhan National Research Foundation (ANRF), Government of India, under Grant Reference No. CRG/2023/003769.

\bibliographystyle{JHEP}
\bibliography{NGLENS.bib}
\end{document}